\documentclass[review]{elsarticle}
\usepackage{geometry}                		% See geometry.pdf to learn the layout options. There are lots.
\usepackage[parfill]{parskip}    		% Activate to begin paragraphs with an empty line rather than an indent
\usepackage{graphicx}				% Use pdf, png, jpg, or epsÂ§ with pdflatex; use eps in DVI mode
\usepackage{float}						% TeX will automatically convert eps --> pdf in pdflatex		
\usepackage{amssymb}
\usepackage{amsmath}
\usepackage{bm}
\usepackage{bbm}
\usepackage{xcolor}
\usepackage{caption,subcaption}
\usepackage{multirow}
\usepackage[ruled,vlined]{algorithm2e}
\usepackage{algorithmic}
\usepackage[numbers]{natbib}%
\usepackage[toc,page]{appendix}
\usepackage{enumitem}
\usepackage{xr}

\journal{Knowledge-based systems}
\begin{document}
\begin{frontmatter}

\title{Diverse personalized recommendations with uncertainty from implicit preference data with the Bayesian Mallows Model}
%\author{Qinghua Liu, Andrew Henry Reiner, Arnoldo Frigessi, Ida Scheel }

\author[mymainaddress]{Qinghua Liu}
\ead{qinghual@math.uio.no}
\author[mysecondaryaddress]{Andrew Henry Reiner}
\ead{a.h.reiner@medisin.uio.no}
\author[mysecondaryaddress,mythirdaddress]{Arnoldo Frigessi}
\cortext[mycorrespondingauthor]{Corresponding author}
\ead{arnoldo.frigessi@medisin.uio.no}
\author[mymainaddress]{Ida Scheel\corref{mycorrespondingauthor}}
\ead{idasch@math.uio.no}

\address[mymainaddress]{Department of Mathematics, University of Oslo, P.O. Box 1053 Blindern, 0316 Oslo, Norway}
\address[mysecondaryaddress]{Oslo Center for Biostatistics and Epidemiology, Oslo University Hospital, Klaus Torg{\aa}rds vei 3,0372 Oslo, Norway}
\address[mythirdaddress]{Oslo Centre for Biostatistics and Epidmiology, University of Oslo, Sognsvannsveien 9, 0372 Oslo, Norway}

						% Activate to display a given date or no date
\begin{abstract}
Clicking data, which exists in abundance and contains objective user preference information, is widely used to produce personalized recommendations in web-based applications. Current popular recommendation algorithms, typically based on matrix factorizations, often have high accuracy and achieve good clickthrough rates. However, diversity of the recommended items, which can greatly enhance user experiences, is often overlooked. Moreover, most algorithms do not produce interpretable uncertainty quantifications of the recommendations. In this work, we propose the Bayesian Mallows for Clicking Data (BMCD) method, which augments clicking data into compatible full ranking vectors by enforcing all the clicked items to be top-ranked. User preferences are learned using a Mallows ranking model. Bayesian inference leads to interpretable uncertainties of each individual recommendation, and we also propose a method to make personalized recommendations based on such uncertainties. With a simulation study and a real life data example, we demonstrate that compared to state-of-the-art matrix factorization, BMCD makes personalized recommendations with similar accuracy, while achieving much higher level of diversity, and producing interpretable and actionable uncertainty estimation. 
\end{abstract}
\end{frontmatter}

\section{Introduction}Personalized recommendations are widely used to help users and customers sort digital information for their purpose. From online streaming services to e-commerce websites, recommender systems can improve business efficiency, sort search results and enhance user experience by providing users a list of personalized, accurate and diverse recommendations. 

Personalized recommendations are based on the users' preference data, which can be explicit feedbacks such as ratings, and implicit feedbacks such as click stream data. Clicking data is easy to collect, exists in great abundance, and often better reflects user preferences compared to ratings. However, the interpretation of clicking data can be challenging, as there is no direct negative feedback from users {\citep{hu2008collaborative}}, and the data naturally exibits high sparsity {\citep{huang2004applying}}. 

The state-of-the-art approach using implicit feedback for personalized recommendation is the Collaborative Filtering for Implicit Data method developed by Hu et al. \cite{hu2008collaborative}. This method is based on matrix factorization (MF). {It is effective and scalable, and is commonly adopted by commercial applications} \cite{gomez2016netflix}. However, there are some drawbacks, in particular, the lack of interpretable uncertainty quantifications: when an item is recommended to a user, the method does not quantify the reliability of the recommendation. In addition, {the collaborative filtering framework} has a tendency to favor the most popular items. While achieving high accuracy, these recommendations can sometimes lack novelty and diversity. {This phenomenon is referred to as the ``diversity-accuracy dilemma''} {\citep{zhou2010solving, liu2012solving, hou2017solving}}. 

We aim at making personalized recommendations with high accuracy and diversity, as well as providing interpretable uncertainty quantifications. In this paper, we introduce the Bayesian Mallows for Clicking Data (BMCD) method by further developing the approach introduced by Vitelli et al. {\citep{vitelli2017probabilistic}}. We assume that users prefer clicked items to unclicked items, and individual clicking data is subsequently augmented to ranking vectors by enforcing the clicked items to be top-ranked. Through a simulation study and an offline testing with a real life dataset provided by the Norwegian Broadcasting Company (NRK), we compare BMCD's recommendation accuracy and diversity measures with the Collaborative Filtering for Implicit Data method. 

In this paper we summarize the Bayesian Mallows Method in Section \ref{sec:model}. We introduce BMCD, and show how we can make personalized recommendations based on posterior probabilities. We briefly summarize the Collaborative Filtering for Implicit Data method in Section \ref{sec:CF}. In Section\ref{sec:eval}, we introduce the evaluation metrics: accuracy and four diversity metrics. In  Section \ref{sec:simulation} we explain the simulation set up and demonstrate how BMCD makes recommendations with uncertainty quantification. In Section \ref{sec:simResults} we present a detailed comparison of BMCD's performance compared to Collaborative Filtering for Implicit Data. In Section \ref{sec:NRK} we apply both methods on the NRK dataset, and compare their performances. Last, a summary and further work are included in Section \ref{sec:summary}.

\section{Related work}

Collaborative filtering \citep{adomavicius2005toward} is a framework utilizing user-item interaction data to make personalized recommendations through borrowing strength across the pool of users and items. 

User-based collaborative filtering is an early method. For a particular user, the basic idea is first to discover other users who  have similar preferences, often measured by cosine similarities or Pearson's correlation coefficent. After such neighbors are identified, recommendations are made based on an aggregation of the neighbors' preferences. User-based collaborative filtering is intuitive and easy to implement, however, it is often limited by the sparsity of the data as well as scalability. Instead, Sarwar et al. \cite{sarwar2001item} proposed an item-based collaborative filtering algorithm. For a given user, her preference of an unknown item is predicted based on the users' past preferences of the $k$ most similar items.

Matrix Factorization (MF) - based collaborative filtering methods are among the most successful {\citep{koren2008factorization}}. {The MF method proposed by Koren et al. {\cite{koren2009matrix}} is developed for a user-item rating matrix. The data matrix $\mathbb{X}$ has dimensions $N \times n$, where $N$ is the number of users and $n$ is the number of items. Each entry $x_{ij}$ is the rating given by a user $j$ to an item $i$, or is empty. Assume that each user $j$ has rated $\leq n$ items.} MF obtains two reduced-dimension matrices $\bm{U}^{N \times L}$ and $\bm{V}^{n \times L}$, with $L < n$, so that their product will be a full matrix $\mathbb{\hat{X}}$ that approximates the original rating matrix $\mathbb{X}$. $\mathbb{\hat{X}}$ predicts, for each user, the ratings of the items that the user has not rated. Hu et al. \cite{hu2008collaborative} extended the method to implicit data. In this paper, BMCD is compared with the method in \cite{hu2008collaborative}{ since this is the widely adopted, state-of-the-art method.} For more details on collaborative filtering, see \cite{koren2015advances}.

%One of the most popular methods for personalized recommendations \textcolor{red}{is Collaborative Filtering (CF). CF is a framework for predicting user preferences by exploiting users' past feedbacks and borrowing strength across the pool of users. Such methods include neighborhood-based approaches (reference) and matrix-factorization (MF) based approaches (references).}

To address the accuracy-diversity dilemma, Zhou et al. {\cite{zhou2010solving}} proposed a graph-based hybrid method. User-item interactions are modeled as a bipartite graph with users and items represented as vertices. The link between a user and an item exists if the user has interacted with the item, and the recommendation process is equivalent to recovering ``lost'' edges. Through weighted linear aggregation {\citep{burke2002hybrid}}, the hybrid method combines a diversity-driven algorithm inspired by the heat diffusion process, and an accuracy-driven algorithm similar to a random-walk process, in order to balance the accuracy-diversity trade-off. This method has achieved improvements in both accuracy and diversity compared to two simple baseline methods, namely, {global ranking (which recommends items according to their overall popularity) and user-based collaborative filtering. However, Zhang et al. \cite{zhang2012auralist} questioned whether multiple diversity objectives can be achieved by such hybrid methods.} Karakaya and Tevfik \cite{karakaya2018effective} introduced a modification of Koren et al. \cite{koren2009matrix}'s MF model for explicit feedback by penalizing popular items to improve diversity. {The method has not been extended to implicit datasets.} 

{Postprocessing of recommendations can help enhance diversity. Antikacioglu et al. \cite{antikacioglu2017post} proposed a bipartite graph-based post-processing method. After a recommendation model is fitted, a score for each user-item pair is obtained, which serve as weights for the user-item edges. The recommendation process is modelled as a maximum-weight bipartite graph matching problem, and diversity is achieved by imposing diversity-related constraints to the optimization, which can be solved using algorithms for minimum cost flow problems. This method could post-process both BMCD and the Collaborative Filtering for Implicit Data Method, but we do not pursue this any further.

\section{Bayesian Mallows for Clicking Data (BMCD) \label{sec:model}}
Consider a dataset of $N$ users and $n$ items $\mathcal{A} = \{A_1, A_2,..., A_n\}$. Suppose first that each user $j$ indicates her preferences with a ranking of all $n$ items, $\bm{R}_j = \{R_{1j},R_{2j},.., R_{nj}\}$, where $R_{ij}$ is the rank assigned to item $i$ by user $j$ , $i = 1, ..., n, j = 1, ..., N$. The Mallows model is a probabilistic model on the space of permutations of $n$ items $\mathcal{P}_n$. In the simplest case, assuming that all users share a common latent consensus $\bm{\rho} \in \mathcal{P}_n$, it has the form of $P(\bm{R}_j = \bm{r}|{\alpha}, \bm{{\rho}}) = {1\over {Z_n(\alpha,\bm{\rho})}} \text{exp} \{-{\alpha \over n}d(\bm{r},\bm{\rho})$\}, where $\alpha$ is a scale parameter, and $d(\bm{r}, \bm{\rho})$ is a distance between $\bm{r}$ and $\bm{\rho}$. Possible choices of distance functions include the footrule distance, the Spearman distance, and the Kendall distance. In this paper, we choose the footrule distance, defined as $d(\bm{r}, \bm{\rho}) = \sum\limits_{i=1}^{n} |{r_i} - {\rho}_i|$, because of its effectiveness {\citep{liu2018model}}. Other distances can also be used. Lastly, $ {Z_n(\alpha,\bm{\rho})} = \sum\limits_{\bm{r}\in \mathcal{P}_n}\text{exp}\{-{\alpha \over n} d(\bm{r}, \bm{\rho})\}$ is the normalizing function. As the footrule distance is a right-invariant distance function, the partition function $Z_n$ is independent of $\bm{\rho}$, and only depends on $\alpha$, hence we denote it as $Z_n(\alpha)$. For $n <50$, $Z_n{(\alpha)}$ has been computed {\citep{vitelli2017probabilistic}}, but is otherwise not analytically computable. When $n \geq 50$, the asymptotic approach  introduced by Mukherjee et al. {\cite{mukherjee2016estimation}} and the importance sampling scheme introduced in {\cite{vitelli2017probabilistic}} are available.

Realistically however, it is uncommon that all users are homogeneous. Assume that the $N$ users are grouped in $C$ clusters, and within each cluster, users share a common latent consensus. For each of the homogenous clusters, we assume a Mallows distribution with parameters $\alpha_c$, $\bm{\rho_c }, c= 1,..., C$. The random variable denoted by $z_j \in \{1, ...,C\}$ assigns user $j$ to cluster $z_j$. Assuming that users' preferences are conditionally independent given the Mallows parameters and their cluster assignments $z_j$, the likelihood function is hence
\begin{equation}
P(\bm{R}_1, ..., \bm{R}_N|\{\alpha_c, \bm{\rho}_c\}_{c=1,...,C}, z_1, ..., z_N) =  \prod_{j=1}^{N} [{Z_n(\alpha_{z_j})}]^{-1}\text{exp}\{-{\alpha_{z_j} \over n}d(\bm{R}_j, \bm{\rho}_{z_j})\}. 
\end{equation}

%Making inference directly on the model parameters via the MLE approach is extremely challenging, as the search space for $\bm{\rho}$ is potentially the space of permutation. 
Vitelli et al. {\cite{vitelli2017probabilistic}} introduce a Bayesian version of this model. The Mallows parameters $\{\alpha_c,\bm{\rho}_c \}_{c=1,...,C}$ are assumed a priori mutually independent. An exponential prior with hyperparamter $\lambda$ is chosen for $\alpha_c, c = 1,..., C$, i.e., $\pi(\alpha_1, .., \alpha_c|\lambda) = \lambda^C\text{exp}\{-\lambda \sum_{c=1}^{C}\alpha_c\}$. For $\bm{\rho}_c, c=1,...,C$, the noninformative uniform prior $\pi(\bm{\rho}_1,..., \bm{\rho}_C) = n!^{-C}$ is chosen. The prior for the cluster assignments $z_j, j=1,...,N$ is $p(z_1,...,z_N|\tau_1,..., \tau_C) = \prod_{j=1}^{N}\tau_{z_j}$, where the probabilities $\tau_1,...,\tau_C$ follow a Dirichlet prior $\pi(\tau_1,...,\tau_C) = \Gamma{(\psi C)}\Gamma({\psi}^{-C})\prod_{c=1}^{C}\tau_c^{\psi-1}$, {$\psi > 0$}. Hyperparameters $\psi$ and $\lambda$ are assumed to be fixed, see \cite{vitelli2017probabilistic} for guidelines.

The posterior distribution of \{$\{\alpha_c, \bm{\rho_c}\}_{c=1, ..., C}$, $z_1, ..., z_N$\} is therefore 

$P(\{\alpha_c, \bm{\rho}_c\}_{c=1,...,C}, z_1,...,z_N|\bm{R}_1,...,\bm{R}_N) \propto\pi(\alpha_1,..., \alpha_C|\lambda)\pi(\bm{\rho}_1,...,\bm{\rho}_C)p(z_1,...,z_N|\tau_1,..,\tau_C)$\\
\begin{equation}
\cdot\pi(\tau_1,...,\tau_C) P(\bm{R}_1,...,\bm{R}_N|\{\alpha_c,\bm{\rho}_c\},z_1,...,z_N).
\end{equation}

We will now extend the Bayesian Mallows model to clicking data. For clicking data, the full ranking of the $n$ items is not available and needs to be inferred from the clicking data. We denote the latent, full  individual ranking vector for user $j$ as $\tilde{\bm{R}}_j$. Suppose that each user $j$ has clicked on a subset of the items  $\mathcal{A}_j \subseteq \mathcal{A}$, with the number of clicks  $|\mathcal{A}_j| = c_j$. % The complement set  $\mathcal{A}_j^c$ indicates the set of items that were not clicked by user $j$, $|\mathcal{{A}}_j^c| = n-c_j$. 
%We use the notation $A_p$ $\succ$ $A_q$ to indicate that item $A_p$ is preferred to item $A_q$. 
It is common to assume that a clicked item is preferred by the user to any other un-clicked item {\citep{joachims2002optimizing}}. For each user $j$, %a set of pairwise preferences can be deducted from the clicks: $\mathcal{B}_j = $ \{$A_p \succ$ $A_q$,  $\forall (p,q)$ s.t. $A_p \in \mathcal{A}_j, A_q \in \mathcal{A}_j^c$\}.
the set of rankings compatible with this assumption is
$\mathcal{S}_j (\mathcal{A}_j) = \{\tilde{\bm{R}}_j \in \mathcal{P}_n \text{ s.t. } \tilde{R}_{ij} < \tilde{R}_{kj} \text{ if } A_i \in \mathcal{A}_j \text{ and } A_k \in \mathcal{A}_j^c \text{ , }\forall i,k, i\neq k\}$.

Given the clicking data, the goal is hence, to sample from the posterior distribution
$$P(\{\alpha_c, \bm{\rho}_c\}_{c=1,...,C},z_1,...,z_N|\mathcal{A}_1,...,\mathcal{A}_N)$$
\begin{equation}
\\=\sum_{\tilde{\bm{R}}_1\in\mathcal{S}_1(\mathcal{A}_1)}...\sum_{\tilde{\bm{R}}_N\in\mathcal{S}_N(\mathcal{A}_N)}P(\{\alpha_c, \bm{\rho}_c\}_{c=1,...,C},z_1,...,z_N, \tilde{\bm{R}}_1,..\tilde{\bm{R}}_N|\mathcal{A}_1,...,\mathcal{A}_N)
\end{equation}

To make inference, we follow a Markov Chain Monte Carlo (MCMC) scheme similar to the one in {\cite{vitelli2017probabilistic}}. Each iteration of the algorithm consists of three major steps: 
\begin{enumerate}[label=(\roman*)]
	\item{\label{step:1}} Update the parameters \{$\alpha_c, \bm{\rho}_c$\} within each cluster $c = 1, .., C$, given the current values of the individual rankings $\tilde{\bm{R}}_j$, and the cluster assignments $z_j$, $j = 1, ..., N$
	\item{\label{step:2}} Re-assign users to clusters based on the current values of the parameters \\$\{\alpha_c, \bm{\rho}_c\}_{c=1, ..., C}$ and the individual ranking vectors $\tilde{\bm{R}}_j$, $j = 1,..., N$
	\item{\label{step:3}} Update $\tilde{\bm{R}}_j$ for each user $j$ given the current values of $z_1, ..., z_N$ and $\{\alpha_c, \bm{\rho}_c\}_{c=1, ..., C}$
\end{enumerate}

As in \cite{vitelli2017probabilistic}, we use a Metropolis-Hasting algorithm for step \ref{step:1}, and a Gibbs sampler for step \ref{step:2}. For step \ref{step:3}, we sample from $P(\tilde{\bm{R}}_1, ..., \tilde{\bm{R}}_N |\{\alpha_c,\bm{\rho}_c\}_{c=1,..., C}, z_1,..., z_N, \mathcal{A}_1, ..., \mathcal{A}_N).$ Given the cluster assignments and the Mallows parameters, the individual rankings are conditionally independent. Therefore, for each user $j$, we can independently sample from the posterior $P(\tilde{\bm{R}}_j |\alpha_{z_j},\bm{\rho}_{z_j}, z_j, \mathcal{A}_j)$ using a Metropolis-Hasting algorithm, where a new $\tilde{\bm{R}_j'}$ for each user must be proposed. One convenient way is to choose two items $i, k$ such that $\{i, k\} \in \mathcal{A}_j$ or $\{i, k\} \in \mathcal{A}_j^c$, and then swap the rankings of the two items for each user $j$. This proposal is symmetric, and each proposed latent full individual ranking vectors $\tilde{\bm{R}}_j'$ is accepted with probability
$\text{min }\{1, \text{exp}[-{{\alpha_{z_j}}\over n} (d(\tilde{\bm{R}}_j', \rho_{z_j}) - d(\tilde{\bm{R}}_j, \rho_{z_j}))     ]\}.$
{Another way of proposing a new $\tilde{\bm{R}}_j'$ is to treat each $\tilde{\bm{R}_j}$ as two parts: the clicked part and the un-clicked part. The ``leap-and-shift'' algorithm in \cite{vitelli2017probabilistic} can then be used separately for the two parts.}

%To compute the normalizing partition function for the footrule distance, when $n \leq 50$, the partition function is calculated analytically, while for $n\geq 50$, the partition function can be approximated using the asymptotic approach  introduced by {\cite{mukherjee2016estimation}} or through importance sampling as in {\cite{vitelli2017probabilistic}}. %Posterior samples of model parameters $\{\alpha_c, 
%\bm{\rho}_c\}_{c=1,..,C}$, cluster assignments $z_1,...,z_N$, and the latent augmented ranking vectors $\tilde{\bm{R}}_j$ for each user can be obtained using an MCMC scheme.

To make personal recommendations, the variables of interest are the latent augmented full ranking $\tilde{\bm{R}}_j$ for each user. For a given user $j$ that has clicked on $c_j$ items, the objective of making $k \geq 1$ recommendations is equivalent to inferring which items are to be ranked as the user's $c_j+1$ -th, ..., $c_j +k$ -th items. We therefore calculate for each user $j$ and each item $i$ the posterior probability to be ranked between $c_j+1$,..., $c_j+k$, which we refer to as the ``next top - $k$" items. That is, we estimate for each user $j$ and each item $i$ 
\begin{equation}\label{equ:tpp}
P_{ij} = P(c_j+1 \leq \tilde{R}_{ij} \leq c_j+k | \mathcal{A}_1, ..., \mathcal{A}_N) = P(\tilde{R}_{ij} \leq c_j+k | \mathcal{A}_1, ..., \mathcal{A}_N).
\end{equation} 

Once estimated, these posterior probabilities are later ranked for each user $j$ in descending order, and the $k$ items with the highest such probabilities are recommended to the user. The estimated top - $k$ probabilities are referred to as the top posterior probabilities (TPP), and the set of $k$ recommended items for user $j$ is denoted as $Rec_{j,k}$. Section {\ref{sec:algo}} in the supplement contains a structured description of our algorithms.

\section{Collaborative Filtering for Implicit Data \label{sec:CF}}
Hu et al. {\cite{hu2008collaborative}} introduced the Collaborative Filtering for Implicit Data method (CF), which extends the classic matrix factorization method. It can be applied to datasets based on implicit user feedbacks, such as clicking data. We now denote the implicit user-item matrix as $\mathbb{X}$. The content of $x_{ij}$ depends on the use case, for example, it can represent the number of times user $i$ has clicked on item $j$. First, a binary matrix $\mathbb{W}$ is introduced by binarizing $\mathbb{X}$ such that $w_{ij}$ is set to 1 if $x_{ij} >0$, and 0 otherwise, {i.e.,} $w_{ij}$ is set to 1 if user $j$ has clicked item $i$, and 0 otherwise. Second, a set of ``confidence" variables $c_{ij}$ is introduced. The rationale behind this variable is that different interactions indicate different levels of certainty that an item is preferred by the user. One choice for $c_{ij}$ is: $c_{ij} = 1+\beta x_{ij}$, $\beta \geq 0$. Finally, the factor matrices are obtained through minimizing the penalized loss function
min$_{\substack{U,V }}$ $\sum\limits_{j,i} c_{ij}(w_{ij} - \bm{u}_j^T\bm{v}_i) + \theta (\sum\limits_{j}||\bm{u}_j||^2+\sum\limits_{i}||\bm{v}_i||^2)$, 
{where both $\bm{u}_j$ and $\bm{v}_i$ are L-dimensional column vectors}. The last term in the loss function is a regularization term and is added to reduce overfitting. The parameters $\beta$, $\theta$, and the reduced dimension of the factor matrices $L$ are determined by cross-validation, while the minimization process is often achieved using algorithms such as alternating least square (ALS){\citep{koren2009matrix}}. In the later sections, the term ``CF'' refers exclusively to the method proposed by Hu et al. {\cite{hu2008collaborative}}, and we use its implementation in Apache Spark {\citep{meng2016mllib}}. BMCD will later be compared with this CF method, in terms of recommendation performances. 

\section{Recommendation evaluation - accuracy and diversity}\label{sec:eval}
In this paper we compare the recommendation performance of BMCD and CF. It is important for a recommendation method to make both accurate and interesting recommendations, and we will assess the two methods in terms of recommendation accuracy, as well as diversity measures. 

To assess recommendation accuracy, the next $k \geq 1$ recommendations are made for each user. In simulations, the recommendations are later compared with the truth. In offline experiments, based on a train-test split of the dataset, accuracy is measured as the percentage of the recommended items that are clicked in the test set. For online experiments, the truth is obtained by experimentation. The drawback of offline experiments compared to online experiments is that the recommendations are not actually given to the users, and hence the truth defined by the test set is not a response to the recommendations. This might be problematic for the recommendation of less popular items, since the users might not even be aware of these items, and hence could not have clicked them in the test set. Offline training-test experimentation is often the only and best alternative for assessing accuracy. 

Despite being an important measure of performance, accuracy is not the only factor that defines successful recommendations {\citep{mcnee2006being, adomavicius2008overcoming}}. User experience can be greatly enhanced when recommendations are diverse, and hence has the potential to be novel and surprising. To assess the diversity of recommendations, we adopt the following four metrics: coverage {\citep{ge2010beyond}}, correct coverage, intra-list similarity {{\citep{ziegler2005improving, zhang2012auralist}} and novelty {{\citep{zhou2010solving}}. 
		%and serendipity. 
		
		\subsection{Coverage}
		Ge et al. {\cite{ge2010beyond}} introduced the metric ``coverage'', defined as
		\begin{center}
			coverage = $\text{\# distinct items recommended to users}\over \text{\# distinct items eligible for recommendation}$,
		\end{center}
		
		the percentage of the  distinct items ever recommended to the users. A recommender system with a high coverage has exploited its pool of items more efficiently, and their users, collectively, are exposed to a wider spectrum of items.
		
		This coverage metric has one major limitation. For a highly inaccurate recommender system, in the extreme case, when recommendations are made randomly, the coverage can be very high while the recommendation accuracy is extremely low. Therefore, we also introduce the ``correct coverage'' metric, defined as:
		
		\begin{center}
			correct coverage = $\text{\# distinct items recommended and clicked by at least one user}\over \text{\# distinct items eligible for recommendation}$
		\end{center}
		
		\subsection{Intra-list similarity}
		
		Ziegler et al. {\cite{ziegler2005improving}} introduced the ``Intra-list similarity'' metric to assess diversity on an individual level. The rationale behind this metric is that, on an individual level, each user tends to prefer recommendations from various categories. A recommendation list that contains items from only one or a few specific categories (for example, a list of only Harry Potter movies), are far less exciting compared to a good mixture of very different items (even for Harry Potter fans). Similarity between two items $a$, $b$ is measured by binary cosine similarity {\citep{zhang2012auralist}} {based on the training data},  defined as 
		\begin{center}
			$CosSim (a,b)$ = ${\text{\# users clicked both a and b}} \over {\sqrt{\text{\# users that clicked }a} \times \sqrt{\text{\# users that clicked }b}}$ , 
		\end{center}  
		
		and the intra-list similarity metric is hence defined as 
		\begin{center}
			
			Intra - list similarity = $1\over N$ $\sum\limits_{{j=1}}^{N}\sum\limits_{a,b \in Rec_{j,k}, b<a} CosSim(a,b)$.
			
		\end{center}
		It is desirable for a recommender to have a low intra-list similarity.
		
		\subsection{Novelty}
		The novelty measure, introduced by Zhou et al. {\cite{zhou2010solving}}, assesses the recommender's ability to recommend items less explored by the users. It is defined as
		\begin{center}
			novelty = $1\over N$ $\sum\limits_{j=1}^{N}\sum\limits_{i\in Rec_{j,k}}$$|\text{log}_2pop_i| \over k$,
		\end{center}
		in which $pop_i$ refers to the popularity of item $i$, in this case, the fraction of all clicks attributed to item $i$ in the training data. A recommender that recommends rare and less popular items, and hence makes novel recommendations, will have a high novelty score. It is desirable to make novel recommendations because a recommendation list consisting of only the most popular items lacks personalization. In addition, the less popular items {are challenging to   be recommended due to lack of data {\citep{adomavicius2008overcoming}}}, and are often valuable to the business {\citep{goldstein2006profiting}}.

\section{Experiment and Results}
In this section, we study the recommendation performances by BMCD and the Apache Spark\citep{meng2016mllib} implementation of Collaborative Filtering(CF) through a simulation study, as well as an offline case study with a dataset provided by the Norwegian Broadcasting Company (NRK). The term CF in the following refers to the Spark implementation of \cite{hu2008collaborative}. Each method will be assessed in terms of recommendation accuracy as well as diversity. 

\subsection{Simulation Study Design\label{sec:simulation}}
In this simulation study, we consider a group of $N = 3000$ users and $ n = 50$ items. The users are partitioned in $C = 3$ equally sized and distinct clusters. The users in each cluster are given full ranking vectors $\bm{R}_j$ sampled from the Mallows model using the sampler in \cite{vitelli2017probabilistic}. For each cluster $c$, the parameters are chosen to be $(\alpha = 3, \bm{\rho_c})$, with $\bm{\rho_1} = \{1,2, .., 50\}$, $\bm{\rho_2} = \{50, 49, ...,1\}$, and $\bm{\rho_3} =$ \{39, 36, 11, 1, 13, 12, 8, 48, 20, 49, 29, 32, 22, 28, 19, 5, 42, 18, 15, 7, 6, 27, 24, 16, 46, 4, 21, 26, 34, 44, 25, 43, 41, 38, 35, 37, 45, 2, 14, 50, 40, 47, 9, 23, 30, 31, 3, 10, 33\}. The 3 consensuses are chosen since they will produce 3 distinct clusters that separate well. Hence, a dense $N$ $\times$ $n$ ranking dataset is obtained,  which will later serve as the ground truth for checking recommendation accuracy, and from which we will build the incomplete clicking dataset.

To simulate clicking data, the full ranking dataset is converted to a binary dataset in the following way. For each user $j = 1, ..., N$, we draw the number of clicks $c_j$ from a truncated Poisson distribution with parameter $\lambda = 5$, truncated to a minimum of 1. Thereafter, the top ranked $c_j$ items are considered ``clicked'', while the rest of the items considered ``unclicked''. In other words, for each user $j$, we obtain $\mathcal{A}_j= \{A_i: R_{ij} \leq c_j\} $.

We generate independently 20 such datasets, and use both CF and BMCD to recommend $k=5$ and $k=10$ items for each user, i.e., to predict for each user $j$ which items are ranked among $c_j+1, ..., c_j+k$. The parameters for CF are determined through 10 - fold cross validation. Although the ground truth dataset is generated from a Mallows model, which can impose some bias towards BMCD in terms of accuracy checking, the dataset is converted to a binary clicking dataset for model fitting. The binarization adds great sparsity to the dataset, and converts ranking vectors into binary vectors, which the Mallows model is not defined for. BMCD's  advantages in inference are considerably reduced due to the binarization. 

To use BMCD, the number of clusters $C$ needs to be determined first. We run {Algorithm \ref{algo:init}}  and {\ref{algo:MCMC}} in the supplementary material with random initialization and varying numbers of clusters $C = 2, 3, ..., 8$. For each value of $C$, we estimate the posterior mean of the sum of within cluster footrule distances (MWCD), defined as
$$
\text{MWCD} = \mathbb{E}[\sum_{c=1}^{C}\sum_{j=1}^Nd(\tilde{\bm{R}}_j, \bm{\rho}_{z_j})|\mathcal{A}_1, ..., \mathcal{A}_N],
$$
by the natural Monte Carlo mean. For each of the 20 simulated datasets, the number of clusters $C$ with the smallest MWCD is chosen. It turns out that $C=3$ is chosen for all runs, except for run numbers 5, 8, and 20, for which $C=4$. Figure {\ref{fig:wcd}} in the {supplementary material Section \ref{sec:simWithinSS}} shows the boxplots of the posterior sum of within cluster distances for 3 selected runs (1, 5, 10). 

The MCMC is run for 1 million iterations, with the first 500000 iterations discarded as burn-in. Similar to the set up in {\cite{vitelli2017probabilistic}}, parameters \{$\alpha_1, ... \alpha_C\}$ are only proposed for update every 10 iterations. The trace plots of $\alpha_c$ for run 1, $c$ = 1, 2, 3, after the burn-in period is included in the {supplementary material Section \ref{sec:convergenceSim}}. Convergence was checked by using multiple starting points of the MCMC chains.
Recommendations of the Bayesian Mallows method are made based on TPP. The recommendation procedure is described in Section {\ref{algo:postproc}} in the supplementary material.

\subsection{Simulation results and discussion \label{sec:simResults}}
\subsubsection{Recommendation accuracy}

After recommendations are made, we refer to the ground truth full ranking vectors $\bm{R}_j$ to check whether the recommended items are truly among each user's next - $k$ items.

Table {\ref{table: sim_mean_next5}} shows the recommendation accuracy using CF and BMCD to predict each user's next $ k = 5$ and $k =10$ items. It can be observed that BMCD makes slightly more accurate recommendations compared to CF in predicting both the next 5 and next 10 items. The accuracy advantage over CF is more significant in the next 10 case. In addition, BMCD's accuracy performance is less varied  than that of CF's.

\begin{table}[h!]
	\centering
	\begin{tabular}{||c c c c c c c c||} 
		\hline
		method&min & 25\% & median & 75\% & max &mean &std dev\\ [0.5ex] 
		\hline\hline
		CF Next-5& 25.02\% & 26.59\% & 27.34\% &27.66\% &28.65\% &27.16\% &0.89\%\\ 
		BMCD Next-5& 26.69\% & 27.54\% & 27.98\% &28.26\% &29.20\% &27.92\% &0.69\%\\
		CF Next-10& 40.54\% & 41.96\% & 42.70\% &43.64\% &44.47\% &42.79\% &1.05\%\\ 
		BMCD Next-10& 43.21\% & 44.33\% & 44.64\% &45.04\% &46.01\%  &44.67\%&0.72\%\\[1ex] 
		\hline
	\end{tabular}
	\caption{Comparisons of accuracies of BMCD and CF, summary of 20 runs}
	\label{table: sim_mean_next5}
\end{table}
We have also discovered that the number of clusters $C$ chosen has little effect on the overall recommendation accuracy, as long as the number of clusters chosen is not too small. As shown in Figure {\ref{fig:sim_C_accuracy}}, the overall recommendation accuracy stablizes after $C \geq 3$.

\begin{figure}[htb]
	\begin{minipage}[t]{.35\textwidth}
		\centering
		\includegraphics[width=\textwidth]{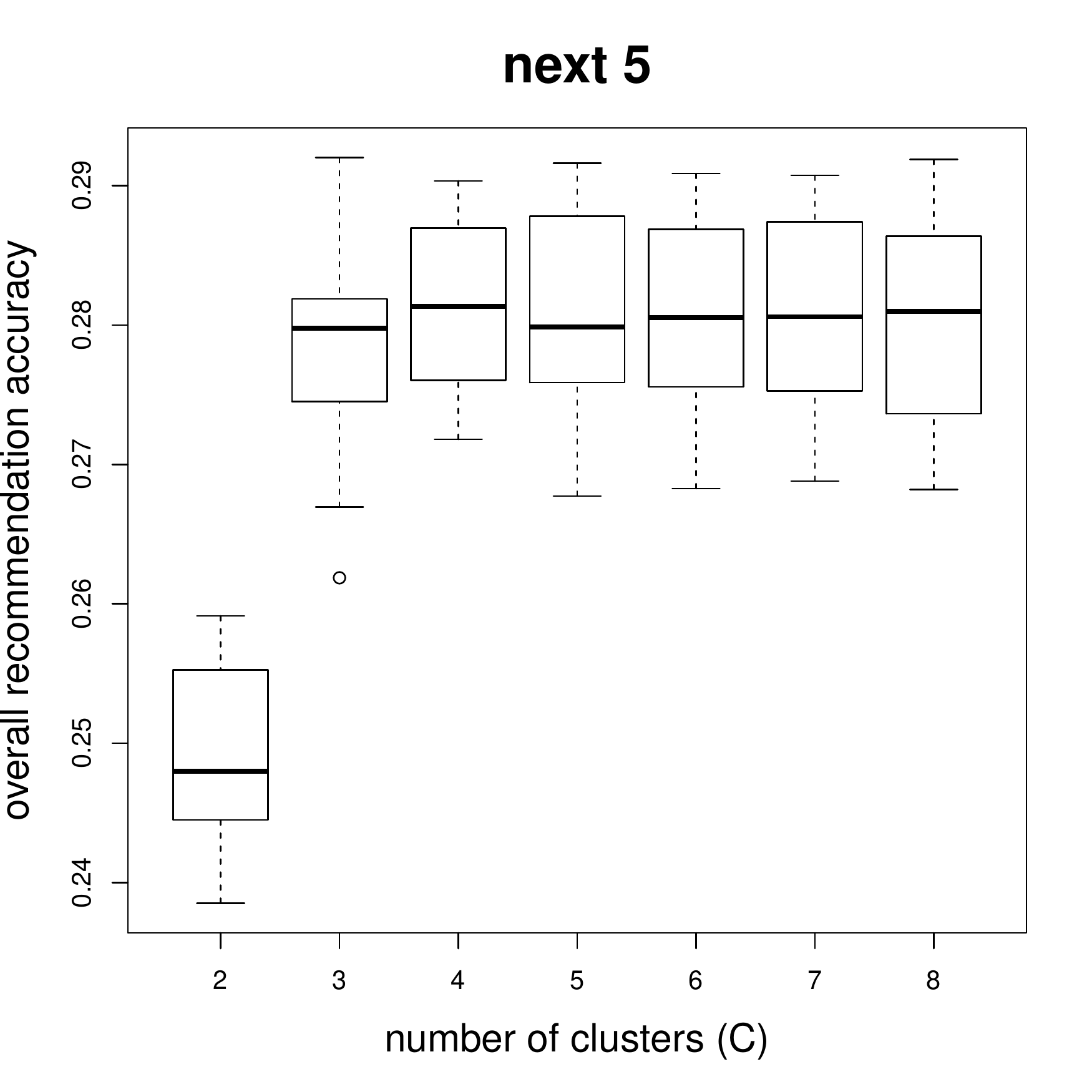}
		\subcaption{Recommend next 5}\label{C_next5}
	\end{minipage}
	\hfill
	\begin{minipage}[t]{.35\textwidth}
		\centering
		\includegraphics[width=\textwidth]{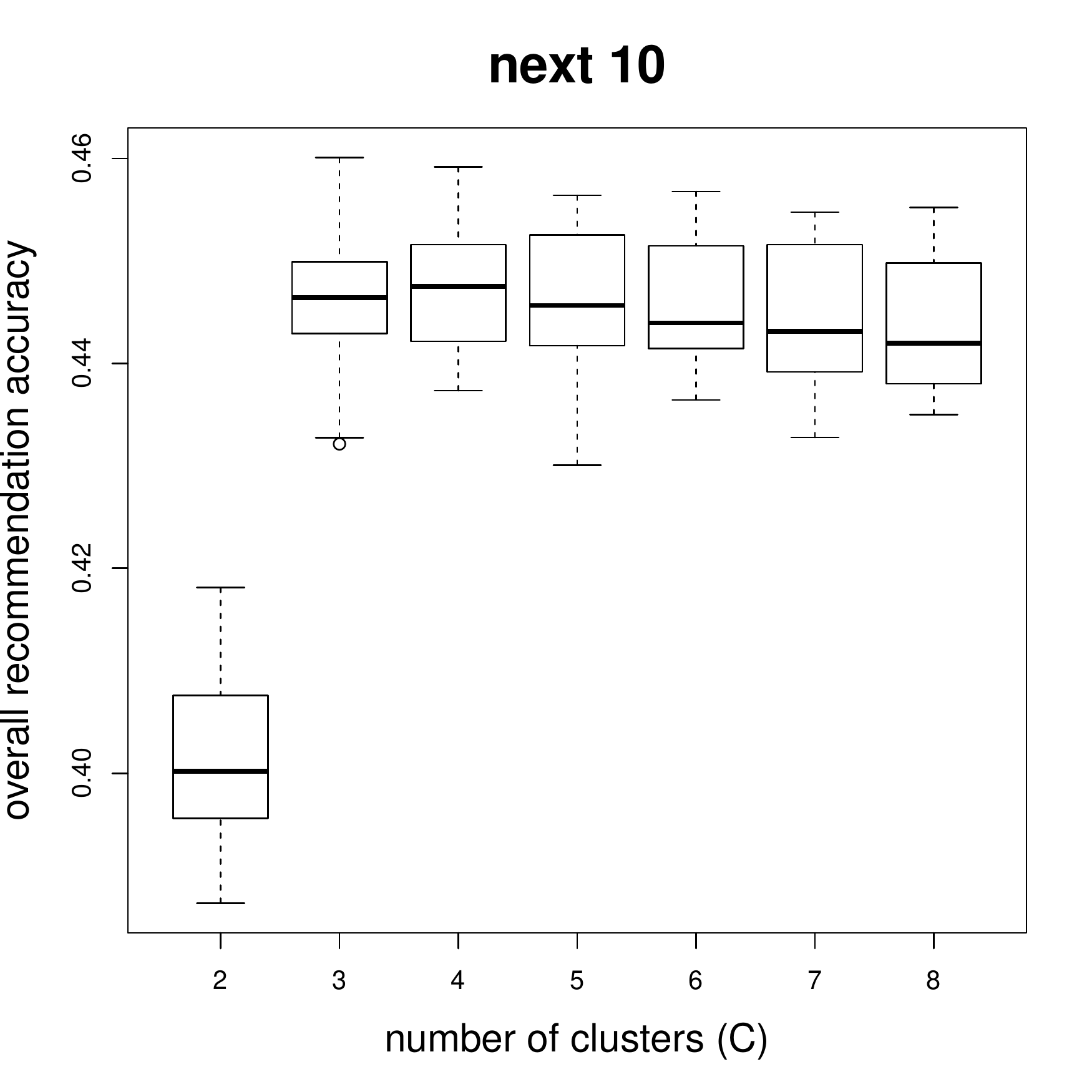}
		\subcaption{Recommend next 10}\label{C_next10}
	\end{minipage}  
	\caption{BMCD's recommendation accuracy vs number of clusters chosen}
	\label{fig:sim_C_accuracy}
\end{figure}

\subsubsection{Recommendation uncertainty quantification}
BMCD also estimates the uncertainty associated with each recommendation through the TPP. Such uncertainties can help assess the reliability of the recommendations by predicting the actual ``hit rates" of the recommended items. For each user $j$ and each recommended item $i$, we can use the binary indicator $t$ to indicate whether the recommended item $i$ is truly among user $j$'s next top - $k$: $t_{ij}= 1$ if ${R}_{ij} \leq c_j+k$, and 0 otherwise.

At the same time, we can bin the TPPs by putting them into $M$ intervals of equal width.  In this case, we choose 0.01 as the bin size. For all TPPs that belong to interval $m$, the associated indicators $t$ are {averaged to }$\bar{t}_m$, indicating the average ``hit rate" of the recommended items associated with the corresponding level of certainty. 

In Figure {{\ref{fig:calibration}}} we plot $\bar{t}_m$ against the binned TPPs for the next $k = 5$ and $k=10$ cases. Run number 10 is shown here, but other runs demonstrate similar trends. The blue dotted $x=y$ line indicates perfect calibration. The red dotted line in the figure represents the percentage of correct recommendations made by CF for this run. From next-5 case, we can clearly observe excellent calibration, especially when the TPPs are in the range between 0.25 and 0.35, where the majority of the recommendations lie within. The uncertainty is not as well calibrated when the TPPs are higher than 0.35 and below 0.25, since there are very few recommendations made with these TPPs. We observe a similar trend in the next - 10 case, also in Figure {{\ref{fig:calibration}}}, with overall higher TPPs, and higher accuracy.

The TPP calculations make it possible for BMCD to identify which recommendations are more reliable than others, because the posterior probabilities are precise and interpretable, and hence can be further exploited. CF on the other hand, produces scores useful for ranking the items but are not easily interpretable. One usage of BMCD's TPPs is introducing a nearly calibrated cut off in order to achieve a higher overall recommendation accuracy. That is to say, we can decide to only make recommendations whose posterior probabilities of being in the next top $k$ has surpassed a threshold and can be expected to be at least the threshold value as hit rate. This will inevitably reduce the number of recommendations made to the users, however, overall accuracy can be expected to be higher. 

\begin{figure}[htb]
	
	\begin{minipage}[t]{.35\textwidth}
		\centering
		\includegraphics[width=\textwidth]{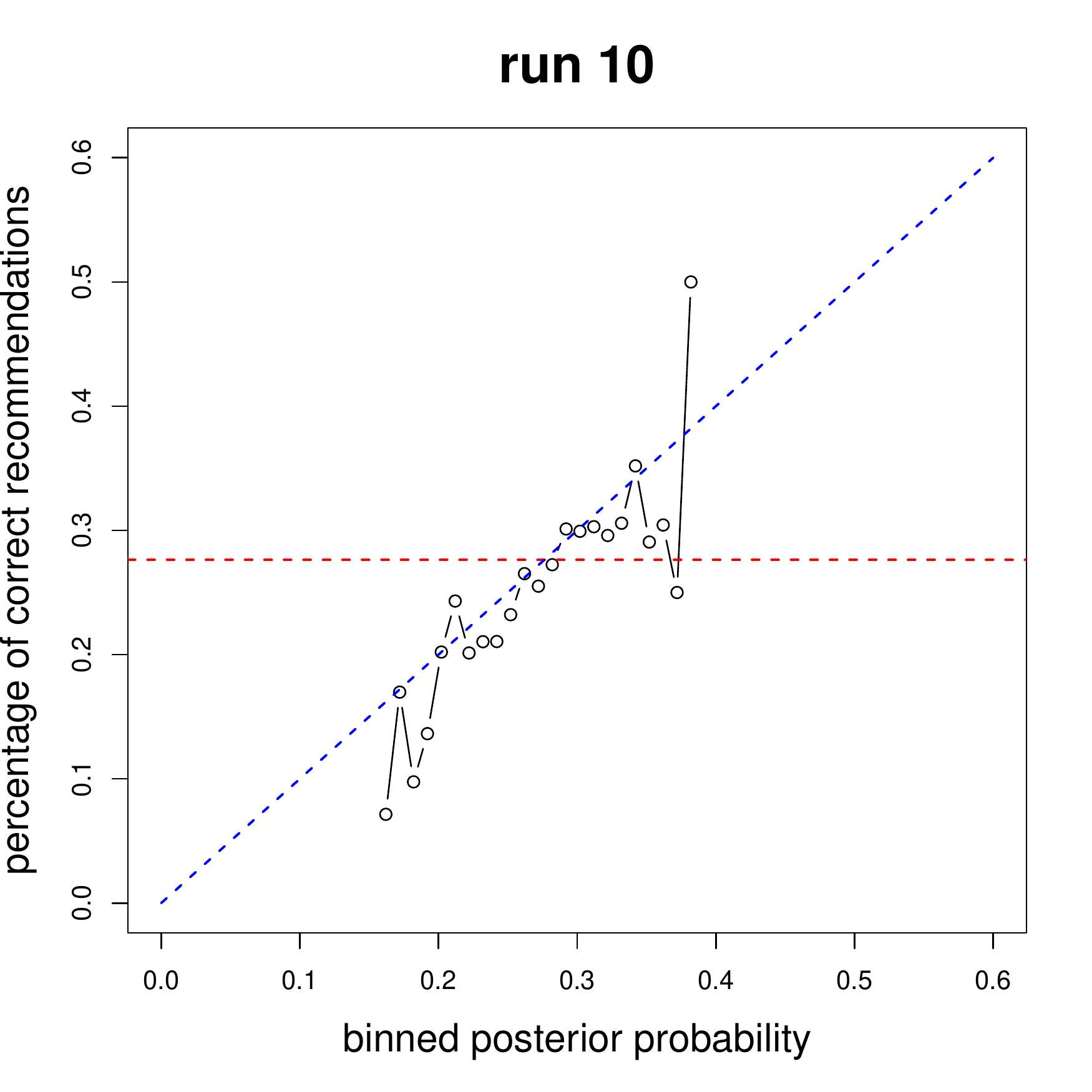}
		\subcaption{Run 10, next-5}\label{calib:3}
	\end{minipage}  
	\hfill
	\begin{minipage}[t]{.35\textwidth}
		\centering
		\includegraphics[width=\textwidth]{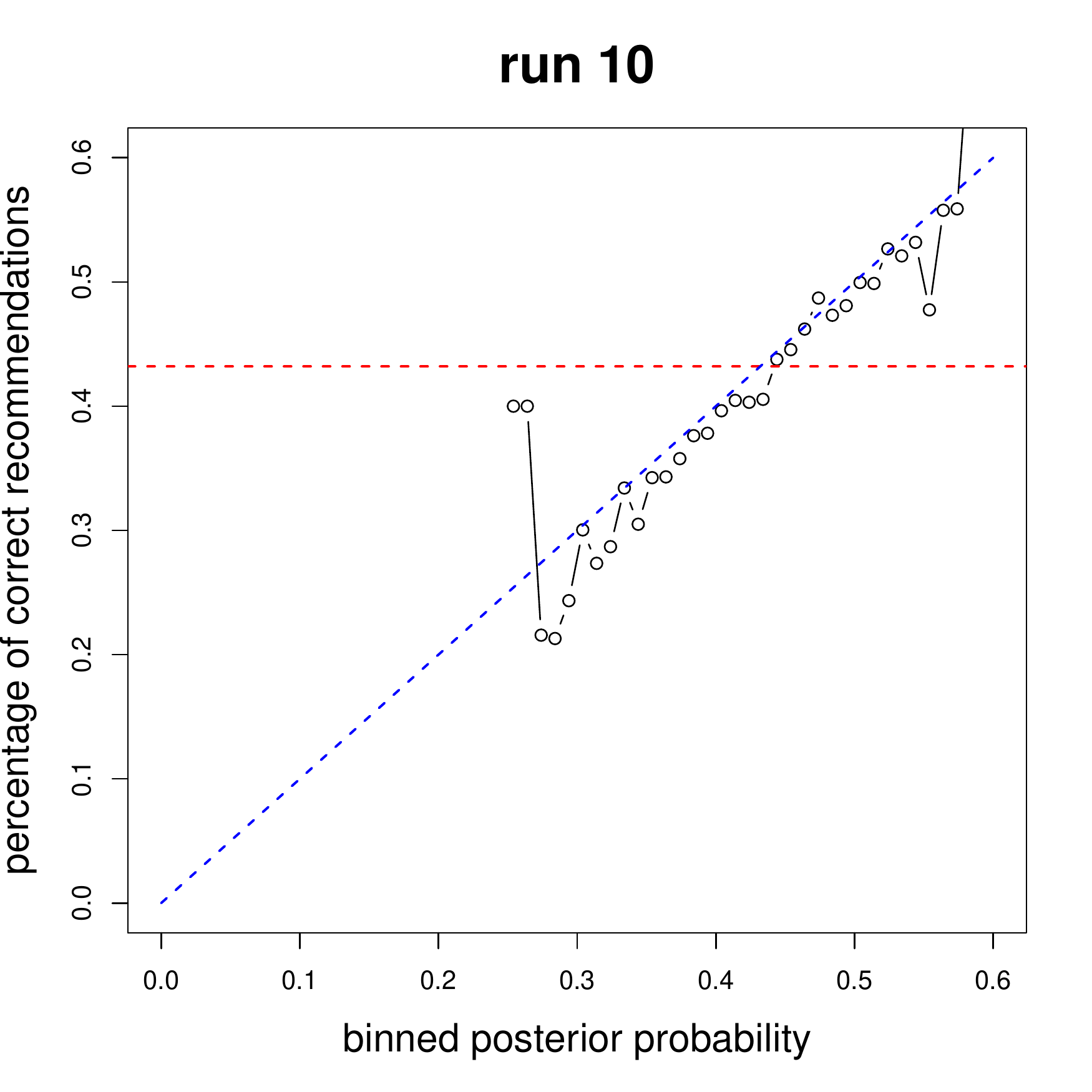}
		\subcaption{Run 10, next-10}\label{calib:6}
	\end{minipage}  
	\caption{Percentage of correct recommendations vs. binned TPPs of one selected run}
	\label{fig:calibration}
\end{figure}

\begin{table}[!htb]
	\begin{minipage}{.47\linewidth}
		\centering
		\caption{Cut off TPPs and the corresponding recommendation accuracies for predicting next 5 items, summary of 20 runs. CF average accuracy: 27.2\%}
		
		\resizebox{\textwidth}{!}{
			\begin{tabular}{||c c c ||} 
				
				\hline
				cut off& number of recommendations & recommendation accuracy\\ [0.5ex] 
				\hline\hline 
				0.10 & 15000$\pm $0 & 28.0 $\pm$ 0.3\% \\ 
				0.15 & 14997$\pm 3.1$ & 28.0 $\pm$ 0.3\% \\ 
				0.20 & 14136$\pm 148.9$ & 28.6 $\pm$ 0.3\% \\ 
				0.25 & 11196$\pm 395.1$ & 29.7 $\pm$ 0.3\% \\ 
				0.30 & 5425$\pm 592.6$ & 31.0 $\pm$ 0.4\% \\ 
				0.35 & 577$\pm 211.1$ & 32.4 $\pm$ 1.5\% \\ 
				0.40 & 145$\pm 9.3$ & 32.5 $\pm$ 8.4\% \\ [1ex] 
				\hline
		\end{tabular}}
		\label{table: cutOff_sim_next5}
	\end{minipage}%
	\hfill
	\begin{minipage}{.47\linewidth}
		\centering
		
		\caption{Cut off TPPs and the corresponding recommendation accuracies for predicting next 10 items, summary of 20 runs. CF average: 42.8\%}
		\resizebox{\textwidth}{!}{
			\begin{tabular}{||c c c ||} 
				\hline
				cut off& number of recommendations & recommendation accuracy\\ [0.5ex] 
				\hline\hline 
				0.20 & 30000$\pm 0.0$ & 44.6 $\pm$ 0.3\% \\ 
				0.25 & 29997$\pm 1.7$ & 44.6 $\pm$ 0.3\% \\ 
				0.30 & 29254$\pm 144.8$ & 45.0 $\pm$ 0.3\% \\ 
				0.35 & 26308$\pm 460.5$ & 46.4 $\pm$ 0.3\% \\ 
				0.40 & 21215$\pm 694.9$ & 48.0 $\pm$ 0.3\% \\
				0.45 & 14468$\pm 762.1$ & 50.0 $\pm$ 0.3\% \\ 
				0.50 & 7104$\pm 800.0$ & 51.0 $\pm$ 0.4\% \\
				0.55 & 1467$\pm 438.4$ & 52.0 $\pm$ 1.2\%\\
				0.60 & 59$\pm 44.0$ & 61.0 $\pm$ 8.2\% \\[1ex] 
				\hline
		\end{tabular}}
		\label{table: cutOff_sim_next10}
	\end{minipage} 
\end{table}

Table {\ref{table: cutOff_sim_next5}} and {\ref{table: cutOff_sim_next10}} show how the recommendation accuracies improve when cut off TPPs are used, for the next - 5 and next - 10 case, respectively. For the next - 5 case, it can be observed from Table {\ref{table: cutOff_sim_next5}} that all of the recommendations made with BMCD have TPPs above 0.1. % and only 20-40 recommendations out of a total of 15000 recommendations are above 0.4. 
Setting a TPP cut off of 0.25 can increase the overall recommendation accuracy by 1.7 percent points compared to not having a cut off, while retaining more than 70\% of the recommendations. Likewise, all TPPs for the next - 10 case are above 0.2, but fewer than 100 recommendations have a TPP of 0.6 or higher. When the cut off TPP is set at 0.45, the number of recommendations are reduced to roughly 50\%, while increasing the overall recommendation accuracy by 5.4 percent points to 50\%.

To summarize, BMCD makes recommendations with similar or slightly higher recommendation accuracies compared to CF in this simulation study. Moreover, the posterior probabilities associated with the recommendations are well calibrated and can be further exploited to assess the reliability of the recommendations. Overall recommendation accuracy can be improved by setting a cut off posterior probability. 

\subsubsection{Diversity}

In this section, we assess both CF and BMCD's abilities to fully exploit the item collection, by making novel and diverse recommendations for each user. We will use the four metrics described in Section {\ref{sec:eval}}.

Table {\ref{tab:simDiv}} summarizes the diversity performances of BMCD and CF. It is desirable to have high values of the coverage, correct coverage and novelty metrics, and a low value of intra-list similarity. We see that recommendations made with BMCD are more diverse and novel compared to CF. BMCD outperforms CF especially on the coverage metric, suggesting that BMCD has stronger ability to discover the less popular items. 
\begin{table}[h!]
	\centering
	\caption{Comparison of diversity performances BMCD and CF}
	\begin{center}
		\resizebox{\textwidth}{!}{
			\begin{tabular}{ccccccc}
				\hline\hline
				metric&min&25\%&median&75\%&max&mean\\
				\hline
				\multicolumn{1}{c}{\multirow{2}{*}{coverage}}
				&CF:0.540 &CF: 0.575 &CF:0.720 &CF: 0.720 &CF:0.760&CF:0.672 \\
				\multicolumn{1}{c}{}
				&BMCD: {0.680}& {BMCD: 0.720}&BMCD: {0.740}& {BMCD: 0.740}& {BMCD: 0.780}& {BMCD: 0.731}\\
				\hline
				\multicolumn{1}{c}{\multirow{2}{*}{corr covg}}
				&CF:0.520 &CF: 0.560 &CF:0.660 &CF: 0.680 &CF:0.700&CF:0.629 \\
				\multicolumn{1}{c}{}
				&BMCD: {0.620}& {BMCD: 0.640}&BMCD: {0.680}& {BMCD: 0.700}& {BMCD: 0.720}& {BMCD: 0.676}\\
				\hline
				\multicolumn{1}{c}{\multirow{1}{*}{intra -list}}
				&CF:1.92 &CF: 2.05 &CF:2.11 &CF: 2.16 &CF:2.23&CF:2.10 \\
				\multicolumn{1}{c}{similarity}
				&BMCD: 1.70& BMCD: 1.78&BMCD:1.91& BMCD:2.06& BMCD: 2.11& BMCD: 1.91\\
				\hline
				\multicolumn{1}{c}{\multirow{2}{*}{novelty}}
				&CF:5.05 &CF: 5.11 &CF:5.18 &CF: 5.20 &CF:5.23&CF:5.16 \\
				\multicolumn{1}{c}{}
				&BMCD: 5.14& BMCD: 5.15&BMCD: 5.29&BMCD: 5.43& BMCD: 5.44& BMCD: 5.29\\
				
				\hline
				%	\multicolumn{1}{c}{\multirow{1}{*}{unserendipity }}
				%	&CF:8.23&CF: 10.3 &CF:11.1 &CF: 12.3 &CF:15.6&CF:11.3 \\
				%		\multicolumn{1}{c}{($\times 10^{-6}$)}
				%		&MLW: 4.10& MLW: 7.19&MLW: 11diversification.4& MLW: 13.6& MLW: 16.2& MLW: 10.3\\
				%			\hline
		\end{tabular}}
	\end{center}
	\label{tab:simDiv}
\end{table}

If we rank all $n$ items according to the number of clicks received by each item (popularity) in the training data in ascending order, and plot the corresponding number of clicks, as shown in Figure {\ref{fig:simFreq}}, it can be observed that the majority of the clicks are received by a small fraction of items. If we define the 20 most clicked items as ``popular'', and the rest of the 30 items as less popular, or ``rare'', we can take a closer look at how often BMCD and CF recommend these ``rare'' items, and how many users have received at least one such rare recommendation. 
\begin{figure}[htb]
	\begin{minipage}[t]{.6\textwidth}
		\begin{minipage}[b]{0.45\textwidth}
			\includegraphics[width=\textwidth]{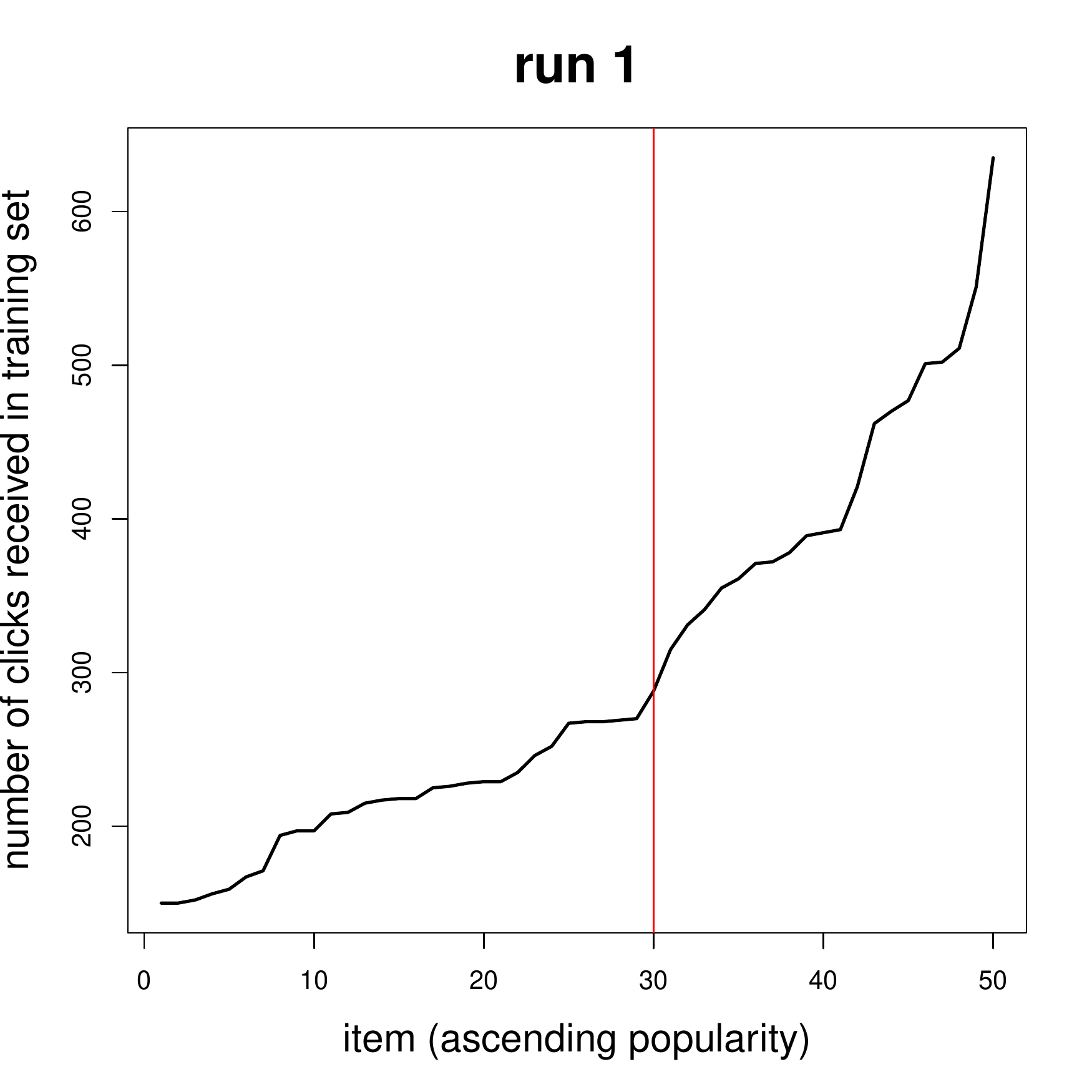}
			\subcaption{Run 1}
		\end{minipage}	
		\hfill
		\begin{minipage}[b]{0.45\textwidth}
			\includegraphics[width=\textwidth]{figures/rePlot_sim/fig3_1}
			\subcaption{Run 5}	
		\end{minipage}	
		\caption{Item popularity of selected runs. The 20 most clicked items are considered as popular}
		\label{fig:simFreq}
	\end{minipage}	
	\hfill
	\begin{minipage}[t]{.35\textwidth}
		\begin{minipage}[b]{\textwidth}
			\includegraphics[width=\textwidth]{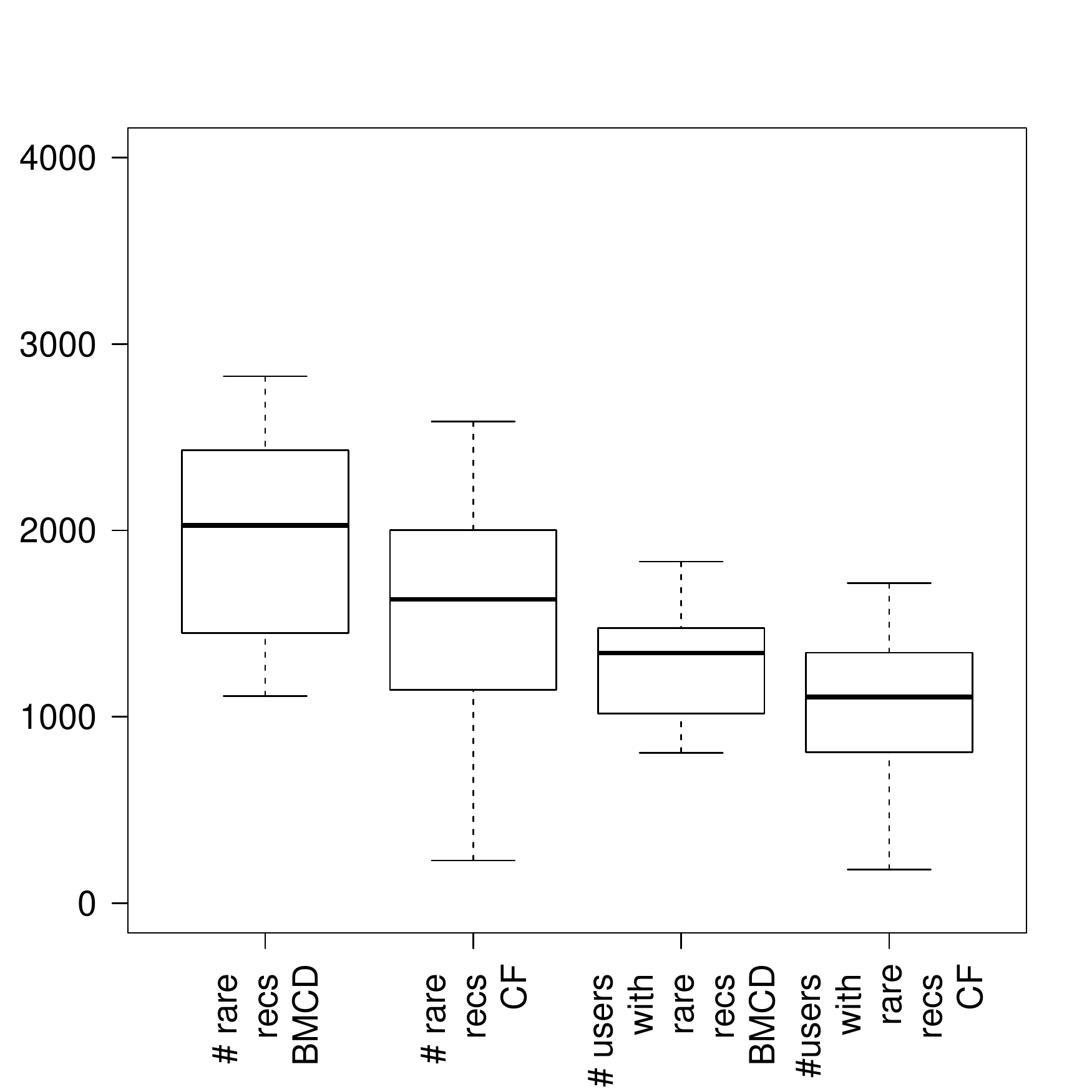}
		\end{minipage}	
		\caption{Comparisons of number of rare item recommendations and number of users with at least 1 recommendations, summary of 20 runs} 
		\label{fig:simRare}
	\end{minipage}	
\end{figure}

From Figure {\ref{fig:simRare}}, it can be seen that BMCD recommends many more rare items, and out of $N = 3000$ users, more than 1000 users have received at least one rare recommendation for all runs, outperforming CF in its ability to explore rare items.

\subsection{Case Study: A clicking dataset from the Norwegian Broadcasting Company (NRK)\label{sec:NRK}}
In this section, we study a dataset containing anonymous log-in users' clicks on movies, TV-series and news programs that are available on the NRK TV website as well as the apps for mobile phones, tablets and other streaming devices such as AppleTV. The data was collected when no personalized recommendation was implemented. We consider only the 200 most popular items. Here for simplification, a whole season of TV series, or a daily news program (which consists of more than one episode), is considered as one single item. Each user - item click is only recorded once, that is, multiple clicks on one item by one user are treated as one click.

We prepare two datasets. Dataset 1 contains all users with at least 13 clicks.  From the clicked items, we make a training set by randomly removing $k$ = 10 clicks per user for prediction purposes. Dataset 2 is a denser subset of dataset 1, which contains users with at least 23 clicks. We make a training set by randomly removing $k$ = 20 clicks per user for prediction purposes. {13 and 23 are chosen to ensure that in the training dataset, each user retains at least 3 clicks}. Table {\ref{table: datasets}} summarizes the two datasets. 
The objective of this study is to make $k$ recommendations for each user, $k = 10$ for dataset 1, and $k = 20$ for dataset 2, using both BMCD, and CF. After recommendations are made, the recommendation accuracy and diversity are studied and compared.

\begin{table}[h!]
	\centering
	\begin{tabular}{||c c c c c c||} 
		\hline
		Dataset& \# of items ($n$) & \# of users ($N$) & min clicks & median clicks &max clicks\\ [0.5ex] 
		\hline\hline 
		1 & 200& 7872 &13 (3) & 18 (8) &103 (93) \\ 
		2 & 200& 2143 &23 (3) & 29 (9) &103 (83) \\ 
		\hline
	\end{tabular}
	\caption{Description of the two NRK datasets. The numbers in parentheses represent the corresponding numbers in the training dataset, after $k$ random clicks per user are removed}
	\label{table: datasets}
\end{table}

\subsubsection{MCMC set up and initialization}
First, the number of clusters $C$ needs to be determined. Alternative to the approach shown in Section {\ref{sec:simulation}}, we used K-means clustering on the NRK binary datasets $\{\mathcal{A}_1,...,\mathcal{A}_N\}$ with different values of $C$, and plot the {within cluster sum of square} against the value of $C$, see Figure \ref{fig:Kmeans} in the supplement. Combining the elbow method and a preference towards a slightly larger number of clusters, which we showed was important in Section \ref{sec:simulation}, $C = 17$ is chosen for dataset 1 and $C = 12$ for dataset 2.

%From {\cref{fig:Kmeans}}, it is possible to notice the change in gradient, but the elbow is not very obvious. However, as we showed in {\cref{sec:simulation}}, it is important that the number of clusters chosen is sufficiently large. Combining the elbow method and a preference towards a slightly larger number of clusters, $C = 17$ is chosen for dataset 1 and $C=12$ for dataset 2.   

While there are many ways of initializing the MCMC, we use the following procedures in order to achieve faster convergence. To initialize the augmented individual ranking vectors ${\tilde{\bm{R}}_{j}}^0$, we first suppose that all users belong to the same cluster, and estimate very roughly a consensus for all users $\bm{\rho}^0$ based on item popularity. We obtain  $\bm{\rho}^0$  by ranking the $n$ items according to the number of clicks each of them has received, and randomize the ties if there are any. Next, we initialize the augmented individual ranking vectors ${\tilde{\bm{R}}_{j}}^0$ based on $\bm{\rho}^0$. First, ${\tilde{\bm{R}}_{j}}^0$ needs to be compatible with the restriction that the clicked items are top-ranked, i.e., 
${\tilde{{R}}_{ij}}^0 \leq c_j $ $\forall A_i \in \mathcal{A}_j$, and ${\tilde{{R}}_{ij}}^0 > c_j $ $\forall A_i \in {\mathcal{A}}_j^c$. Second, while satisfying this restriction, we want to inherit the pairwise comparisons represented in the group consensus $\bm{\rho}^0$. That is to say, for each user $j$,
\begin{center}
	$\forall p,q \in \mathcal{A}_j$ and $\forall a,b \in \mathcal{A}_j^c$ \\	
	${\tilde{R}_{pj}}^0 < {\tilde{R}_{qj}}^0 \leq c_j$,   if ${\rho}^0_p < {\rho}^0_q$	\\
	$c_j < {\tilde{R}_{aj}}^0 < {\tilde{R}_{bj}}^0$,  if ${\rho}^0_a < {\rho}^0_b$. 	\end{center}
For example, if we have a 5 - item set \{A, B, C, D, E\}, $\bm{\rho}^0 = \{1, 2, 3, 4, 5\}$, and user $j$ has clicked on item A, C, and E, the initialization of the augmented vector ${\tilde{\bm{R}}_{j}}^0$ is therefore, \{1, 4, 2, 5, 3\}. This initialization speeds up the MCMC convergence significantly.

The cluster assignment $z_j$ for $j=1, ..., N$, is initialized randomly. Within each cluster $c$, $\bm{\rho}^0_c$ is initialized in a similar manner as $\bm{\rho}^0$, however, the item popularity is calculated only based on the clicks by the users that belong to cluster $c$. The parameters \{$\alpha_c^0\}_{c = 1, ..., C}$ are initialized as $a_1^0 = ... = a_C^0 = 3$, other values can also be chosen.

We run the MCMC for 5 million iterations and 7 million iterations, for dataset 1 and dataset 2, respectively. It takes longer for dataset 2 to reach convergence, presumably since there are more users that swing between different clusters. Only the last 1 million iterations are used for subsequent analyses. The MCMC is thinned at every 100 iterations while  \{$\alpha_1, ... \alpha_C\}$ is proposed every 10 iterations.The trace plots of  \{$\alpha_1, ... \alpha_C\}$ after the burn-in period are shown in Figure {\ref{fig:NRK_alpha}} in the supplement. 
%For dataset 1, most users are concentrated in cluster 1 with a very high value of $\alpha_1$, and as shown in the top-left figure, the distribution of $\alpha_1$ is more peaked compared to the other clusters. For dataset 2, cluster 6 and cluster 12 include more than 50\% of the users. 

\subsubsection{Recommendation accuracy}
Table {\ref{table: accuracy }} shows the overall recommendation accuracy for predicting the next - $k$ items for both datasets using BMCD and CF respectively.
\begin{table}[h!]
	\centering
	\begin{tabular}{||c c c ||} 
		\hline
		Method& Dataset 1 (next 10) & Dataset 2 (next 20) \\ [0.5ex] 
		\hline\hline 
		BMCD &{26.4\%}& {35.0}\% \\ 
		CF & 29.9 \%& 34.9\% \\ 
		\hline
	\end{tabular}
	\caption{Comparison of next-$k$ recommendation accuracies for the NRK datasets}
	\label{table: accuracy }
\end{table}
It can be observed that  CF in this case outperforms BMCD in terms of accuracy for dataset 1, while for dataset 2, the two methods' accuracies are almost identical. The NRK dataset is collected when no personalized recommendation is rolled out. In this situation, all users' clicks are quite concentrated on the popular items. As will be discussed in more detail in Section {\ref{sec:diversity_NRK}}, BMCD's tendency to recommend a more diverse set of items and the inclusion of less popular items, compared to CF, presumably contributes to the slightly inferior recommendation accuracy for dataset 1.

\subsubsection{Uncertainty quantification of BMCD recommendations}
\subsubsection{Uncertainty quantification of BMCD recommendations}

\begin{figure}[htb]
	\begin{minipage}[t]{.35\textwidth}
		\centering
		\includegraphics[width=\textwidth]{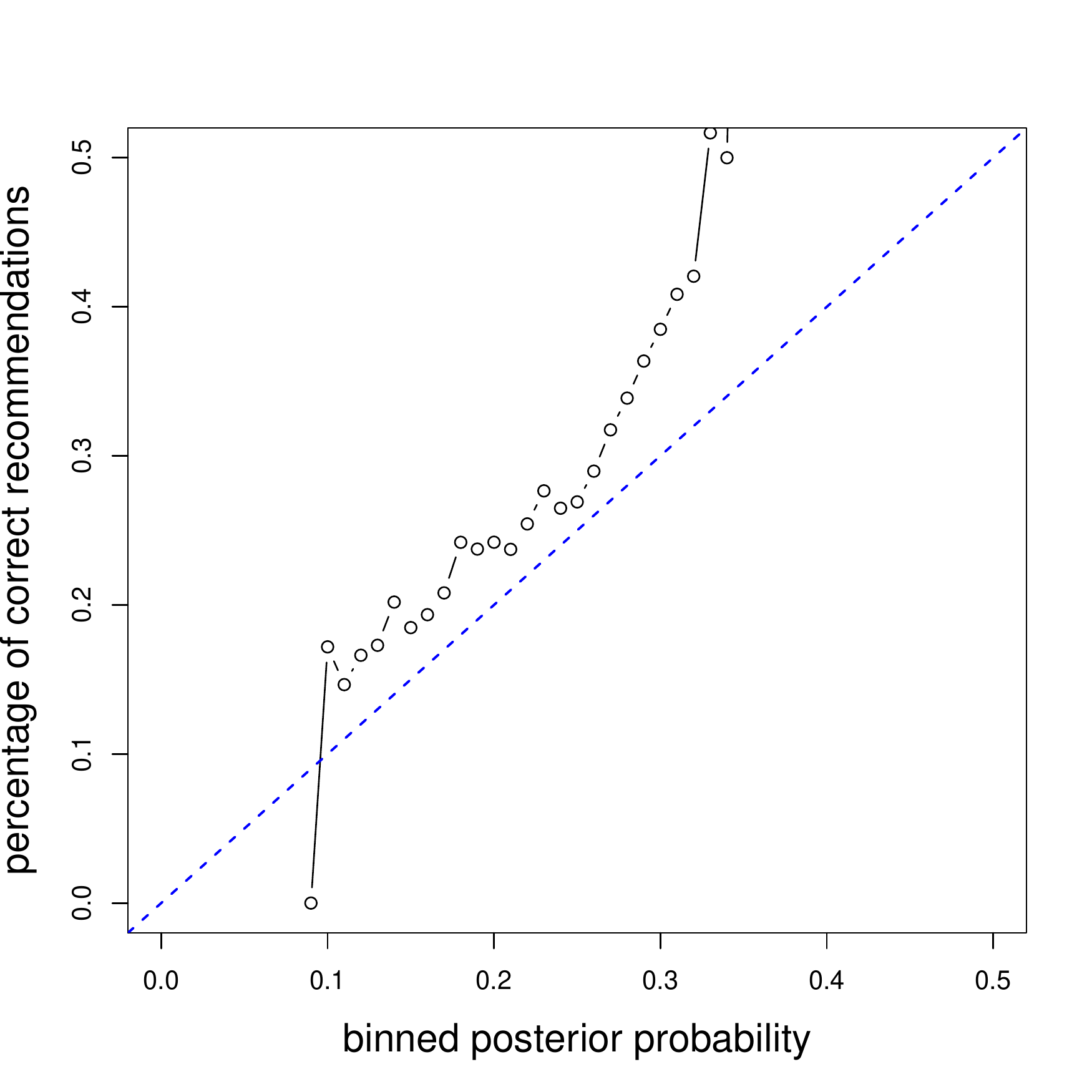}
		\subcaption{Dataset 1}\label{nrk10_calib}
	\end{minipage}
	\hfill
	\begin{minipage}[t]{.35\textwidth}
		\centering
		\includegraphics[width=\textwidth]{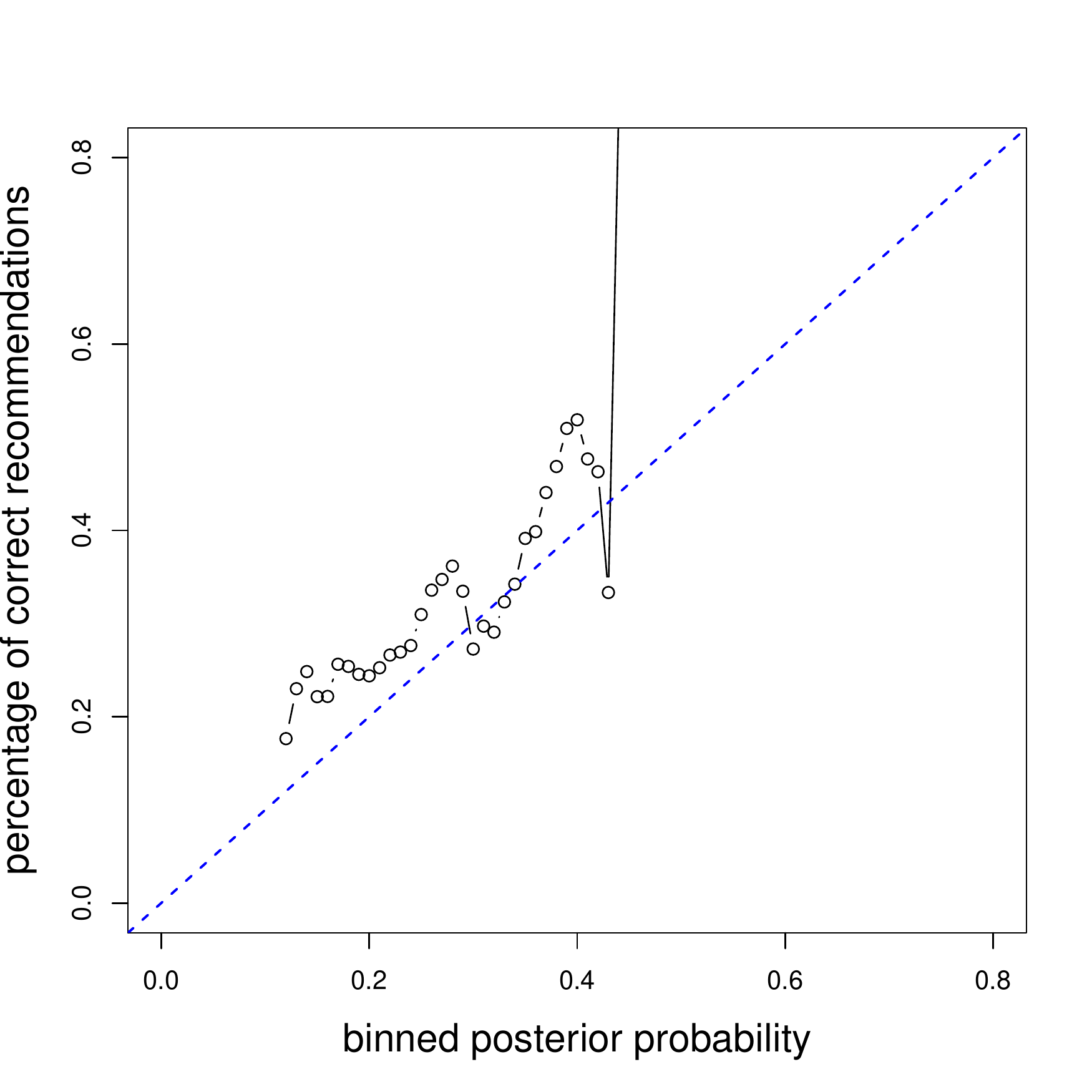}
		\subcaption{Dataset 2}\label{nrk20_calib}
	\end{minipage}  
	
	\caption{Recommendation accuracies vs. binned TPPs for BMCD}
	\label{fig:NRK_calib}
\end{figure}

%\textbf{Figure \textcolor{red}{x}} shows the recommendation frequency plotted against the discretized posterior probability. Within each bin, we show the number of recommendations that are correctly made and wrongly made. It can be observed that most recommendations are associated with a recommendation posterior probability between \textcolor{red}{0.3 and 0.3x}. 

Similar to Figure {\ref{fig:calibration}}, Figure {\ref{fig:NRK_calib}} shows the recommendation accuracy plotted against the binned {TPPs}. A clearly increasing trend can be observed. BMCD in this case, tends to underestimate the certainty of each recommendation made,{ or in other words, the TPPs estimated are slightly lower than the actual hit rates of the recommendations} as the blue line is slightly above the dotted line. This can be explained by a slight misfit of the Mallows model.We can exploit the uncertainty to identify reliable recommendations as well as introducing cut off TPPs to improve overall accuracy of BMCD. }

\begin{figure}[htb]
\begin{minipage}[t]{.35\textwidth}
	\centering
	\includegraphics[width=\textwidth]{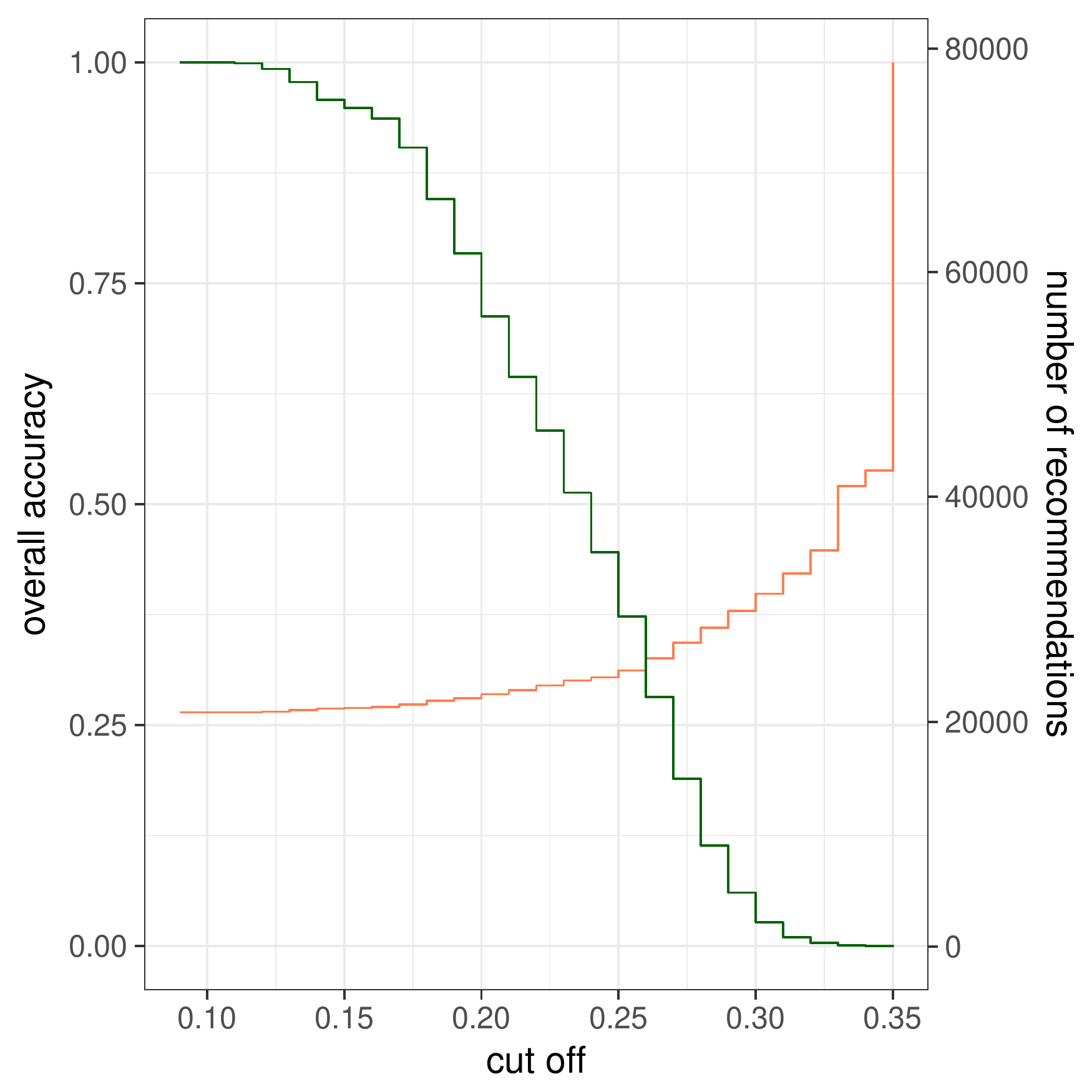}
	\subcaption{Dataset 1}\label{nrk10_cutoff}
\end{minipage}
\hfill
\begin{minipage}[t]{.35\textwidth}
	\centering
	\includegraphics[width=\textwidth]{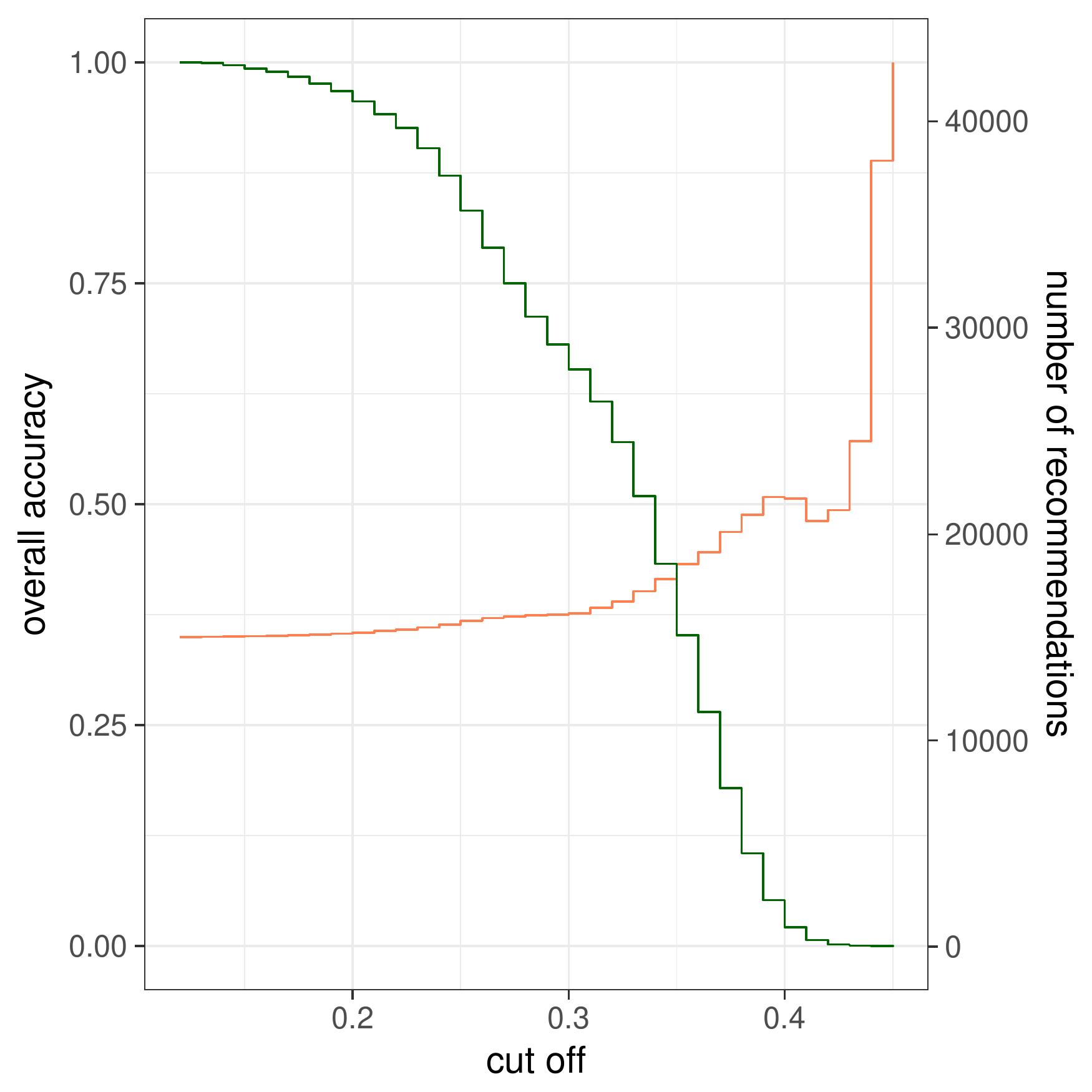}
	\subcaption{Dataset 2}\label{nrk20_cutoff}
\end{minipage}  

\caption{Overall recommendation accuracies and number of recommendations performed vs. cut off posterior probabilities, green line: number of recommendations, red line: overall recommendation accuracies. Y-axis on the left: overall recommendation accuracies. Y-axis on the right: number of recommendations}
\label{fig:NRK_cutoff}
\end{figure}

We see from Figure {\ref{fig:NRK_cutoff}} that, as the cut off TPP increases, the number of recommendations strictly decreases while the overall recommendation accuracy improves. For dataset 1, when the cut off posterior probability is 0.23 or above, BMCD's overall recommendation accuracy exceeds $30\%$, making it identical to CF, while retaining 60\% of the recommendations.

\subsection{Diversity}\label{sec:diversity_NRK}

\subsubsection{Coverage}
The coverage metric is especially important for NRK. As a national broadcaster, NRK has a large collection of valuable historical contents and non-mainstream programs that may be rarely discovered by its users; however, theseprograms have high quality and should be promoted. 
\begin{table}[h!]
	\centering
	\caption{Comparisons of coverage and correct coverage}
	\begin{center}
		\begin{tabular}{ccc}
			\hline\hline
			Dataset&coverage&corr coverage\\
			\hline
			\multicolumn{1}{c}{\multirow{2}{*}{Dataset 1}}&CF:0.61 (122/200) &CF: 0.545 (109/200)\\
			\multicolumn{1}{c}{}&BMCD: {0.865(173/200) }& {BMCD: 0.720(144/200)}\\
			\hline
			\multicolumn{1}{c}{\multirow{2}{*}{Dataset 2}}&CF: 0.38 ($76/ 200$)  &CF: 0.280 (56/200)\\
			\multicolumn{1}{c}{}&BMCD: 0.82 (164/200) &BMCD: 0.685(137/200)\\
			\hline
			
		\end{tabular}
	\end{center}
	\label{tab:coverage}
\end{table}

Table {\ref{tab:coverage}} summarizes the comparisons of coverage and correct coverage of BMCD and CF. Both methods cover a broader range of items for dataset 1, as the dataset contains more users, leading to more diverse preferences. Dataset 2 is a more difficult scenario where the users' preferences are more homogeneous, and it is therefore more challenging to make diverse recommendations. It is clear that BMCD outperforms CF in terms of coverage, and the advantage is especially significant for dataset 2. This suggests that, consistent with the simulation, CF tends to recommend more popular items while BMCD has a stronger ability to explore the rare items. In addition, BMCD does not sacrifice much accuracy for diversity, as it also outperforms CF in the correct coverage metric. 

Figure {\ref{fig:coverage}} shows the recommendation frequency of the items being recommended. On the x-axis, the items are ranked according to their popularity in ascending order. For both datasets, CF's recommendations are much more concentrated on the more popular items. BMCD in comparison, recommends much fewer popular items compared to CF.

\begin{figure}[h!]
	
	\begin{minipage}[t]{.23\textwidth}
		\centering
		\includegraphics[width=\textwidth]{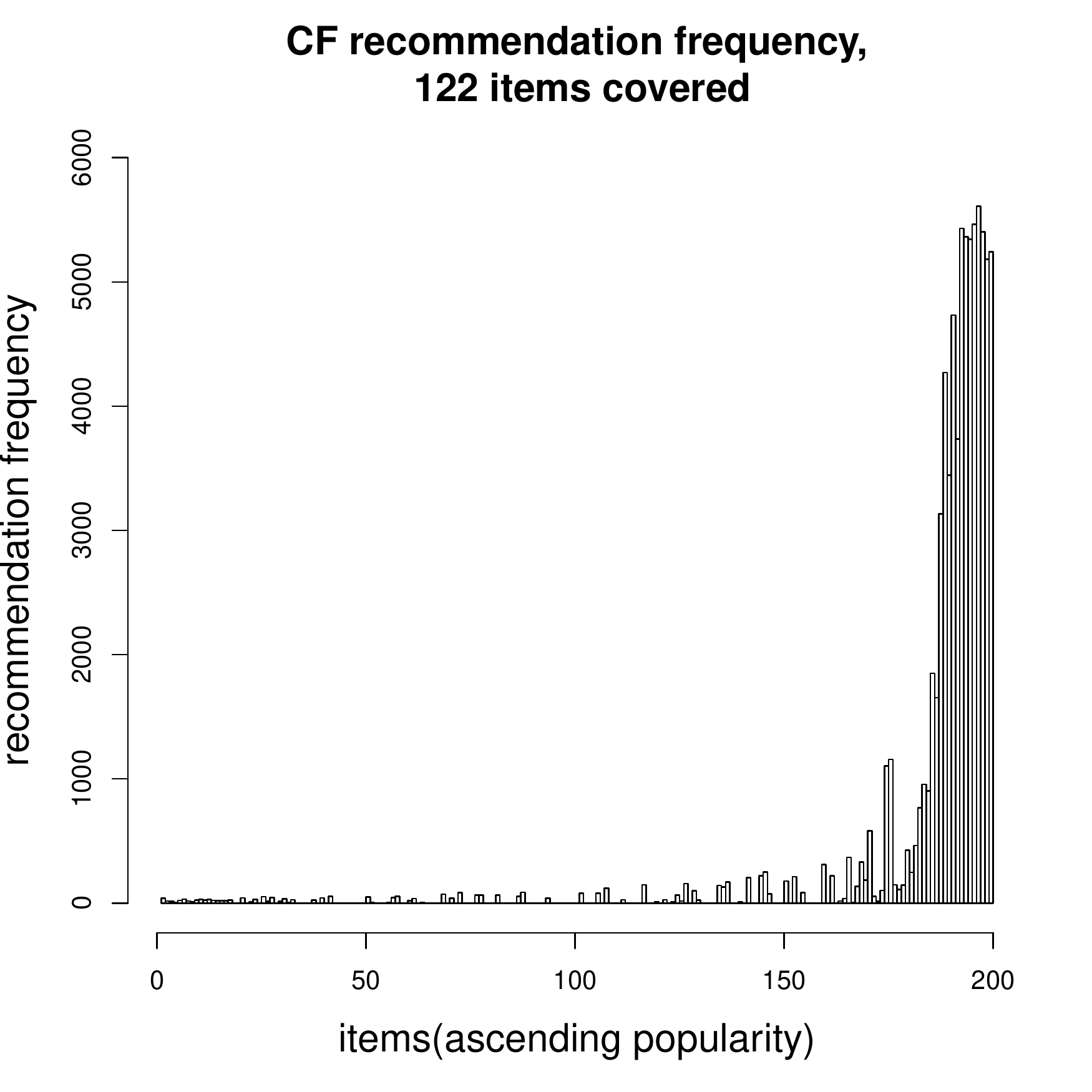}
		\subcaption{Dataset 1, CF}\label{nrk10_corrCovCF}
	\end{minipage}
	\begin{minipage}[t]{.23\textwidth}
		\centering
		\includegraphics[width=\textwidth]{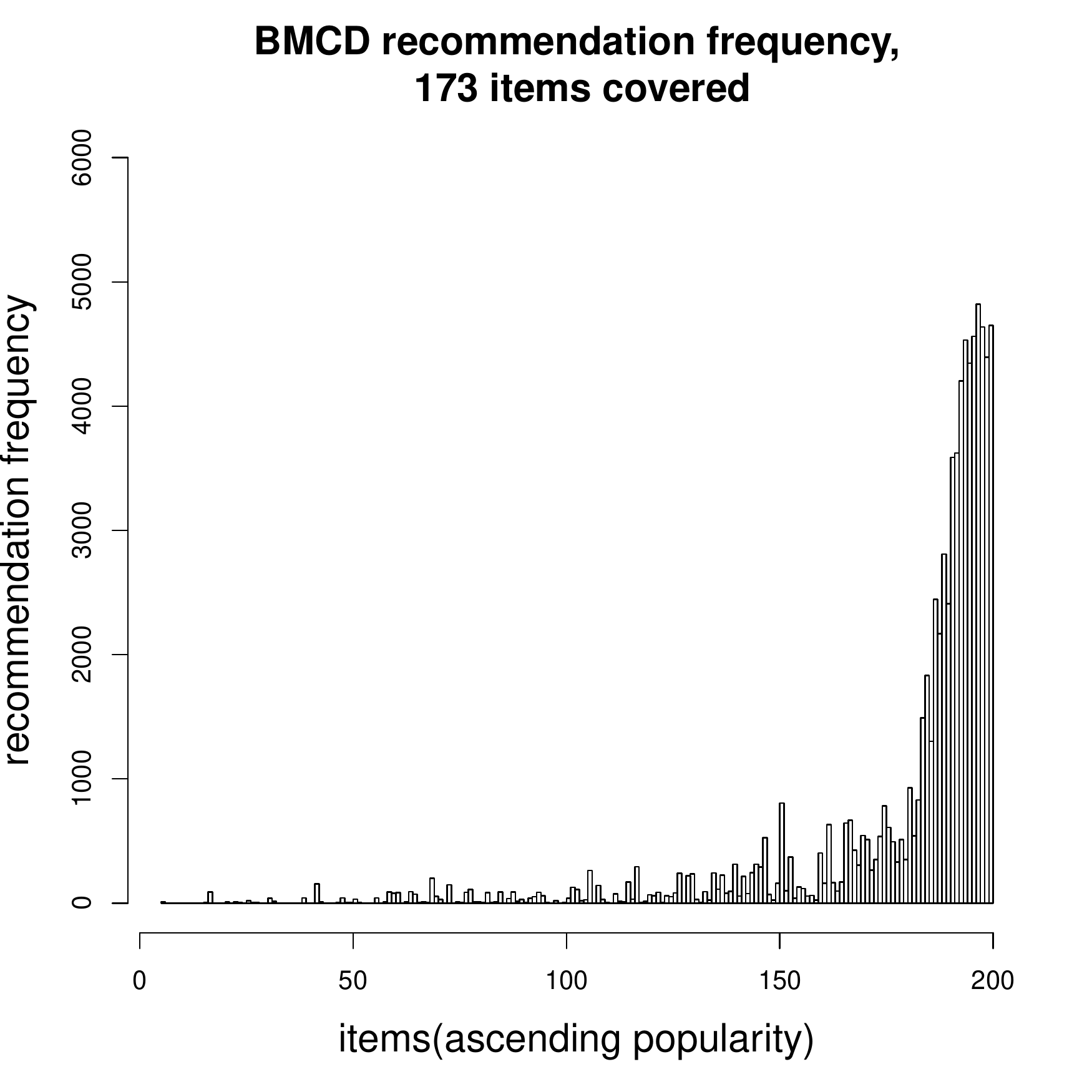}
		\subcaption{Dataset 1, BMCD }\label{nrk10_corrCovML}
	\end{minipage}  
	\begin{minipage}[t]{.23\textwidth}
		\centering
		\includegraphics[width=\textwidth]{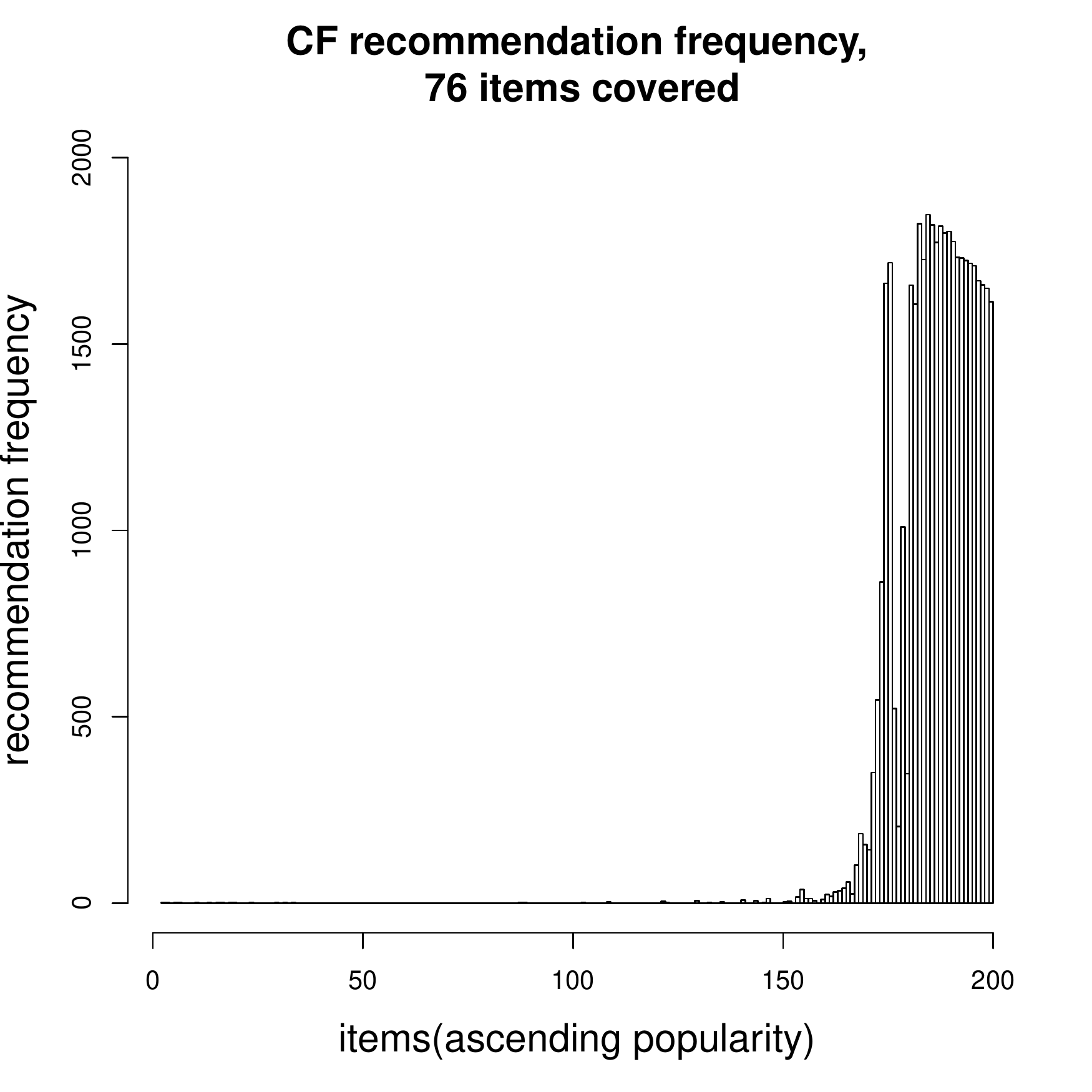}
		\subcaption{Dataset 2, CF}\label{nrk20_corrCovCF}
	\end{minipage}
	\hfill
	\begin{minipage}[t]{.23\textwidth}
		\centering
		\includegraphics[width=\textwidth]{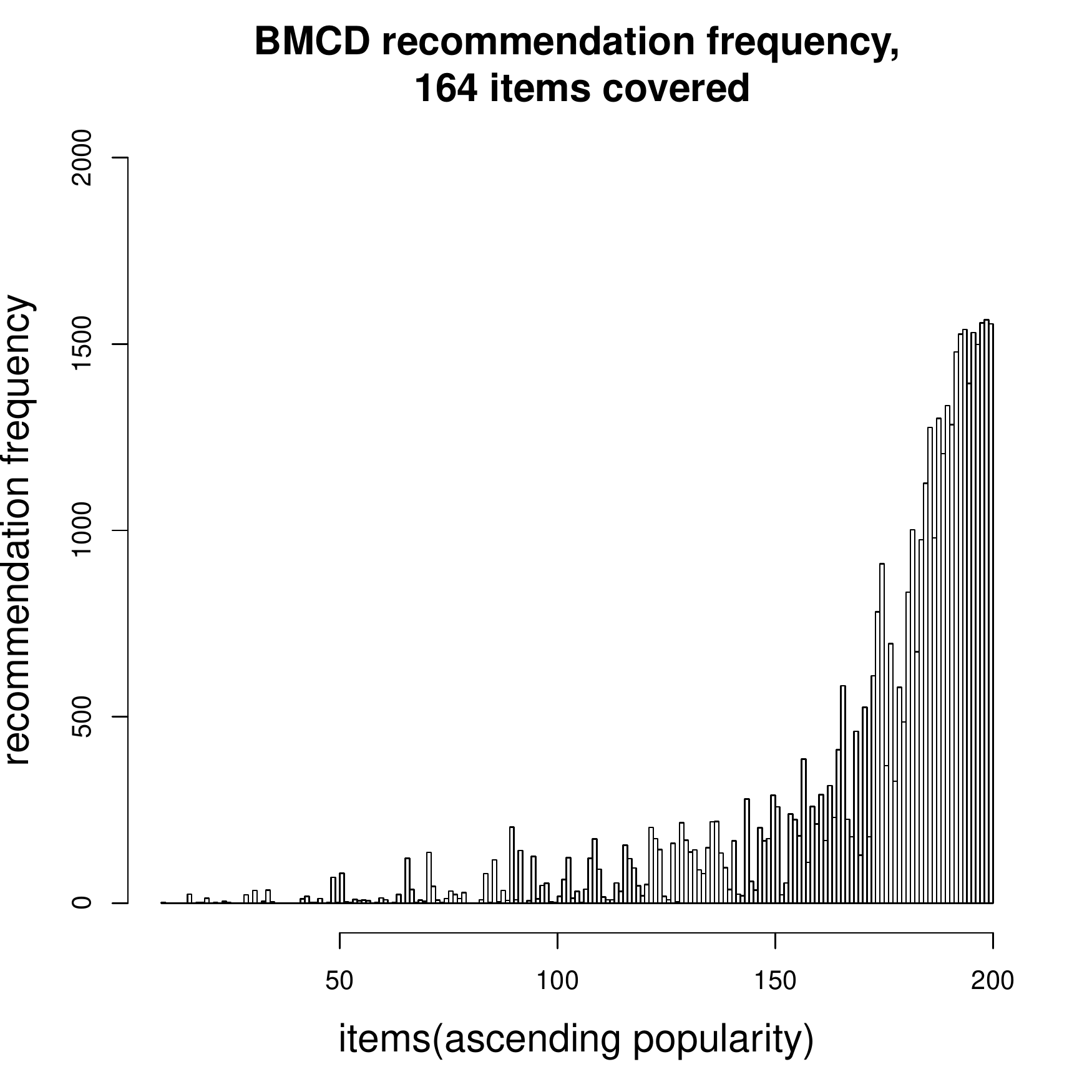}
		\subcaption{Dataset 2, BMCD }\label{nrk20_corrCovML}
	\end{minipage}  
	
	\caption{Histogram of correct recommendations. X-axis: item labels, arranged according to ascending item popularity (measured by the number of clicks received in the training set). Y-axis: recommendation frequency}
	\label{fig:coverage}
\end{figure}

To give a clearer definition of ``popular'' items, Figure {\ref{fig:itemFreq}} shows the number of clicks received by each item in the training dataset, with the x-axis arranged from the least clicked item to the most clicked item. In both datasets, most of the clicks are attributed to roughly the 40 most popular items, which we define as ``popular'' items, while the rest we defined as ``rare'' items. Based on this definition, the number of ``rare'' items recommended by BMCD and CF, and the number of users receiving at at least one ``rare'' recommendations are shown in Table {\ref{table:rareRecs}}. It clearly shows that BMCD makes significantly more recommendations that are less popular compared to CF. In particular, for dataset 1, more than 12.8\% of all recommendations made with BMCD involves rare items and more than 40\% of all users receive at least 1 rare recommendation. For CF, only 5.6\% of all recommendations are rare, while 11.8\% of of the users receive 1 or more rare recommendations. The contrast between CF and BMCD is even more obvious for dataset 2, where only 4\% of all recommendations made with CF involves rare items, compared to BMCD's 20.5\%. Given that CF's and BMCD's recommendation accuracies are similar in  this case, and that BMCD makes more rare recommendations, it follows that BMCD also has a higher recommendation accuracy when recommending popular items compared to CF.

\begin{figure}[h!]
	
	\begin{minipage}[t]{.35\textwidth}
		\centering
		\includegraphics[width=\textwidth]{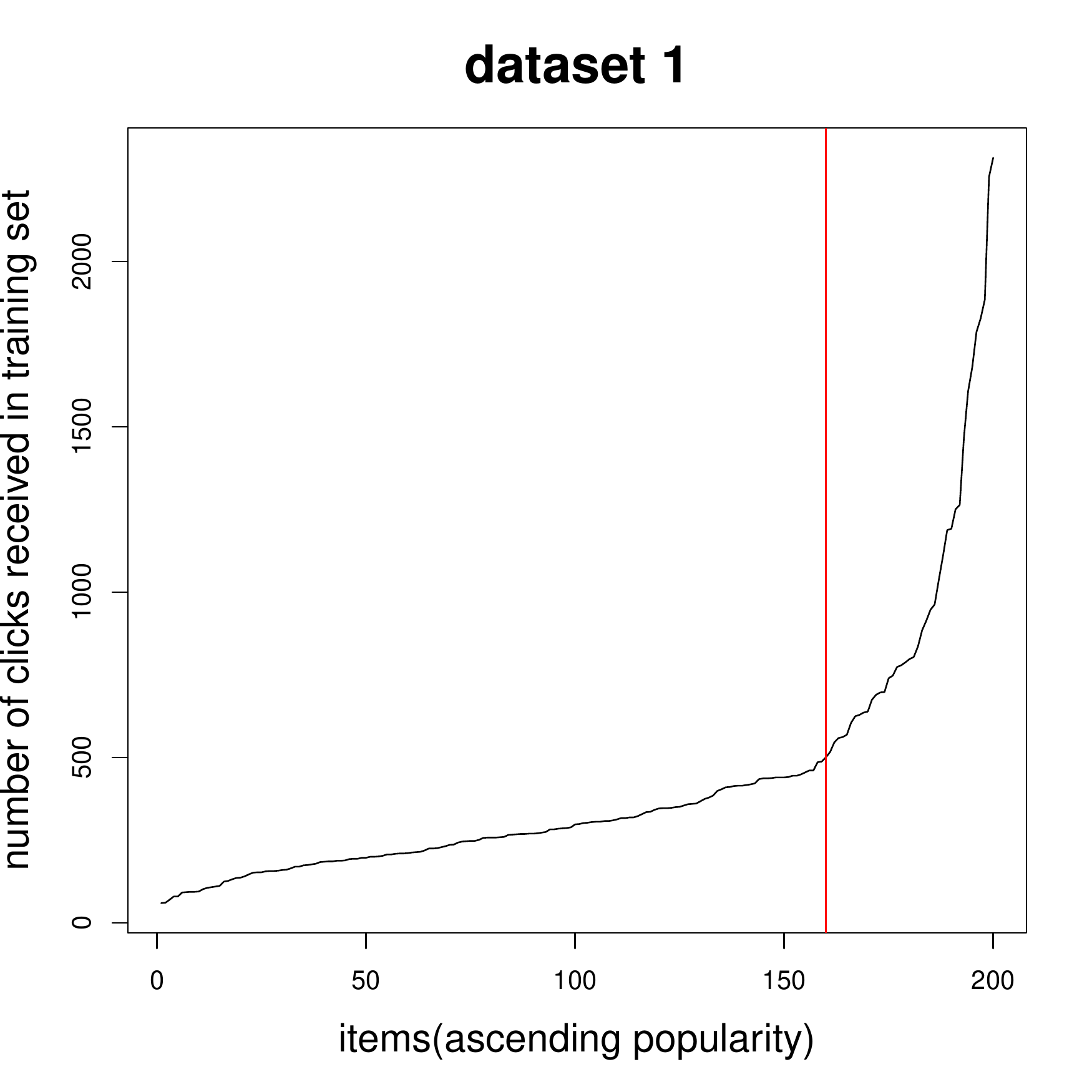}
		\subcaption{Dataset 1}\label{itemFreq10}
	\end{minipage}
	\hfill
	\begin{minipage}[t]{.35\textwidth}
		\centering
		\includegraphics[width=\textwidth]{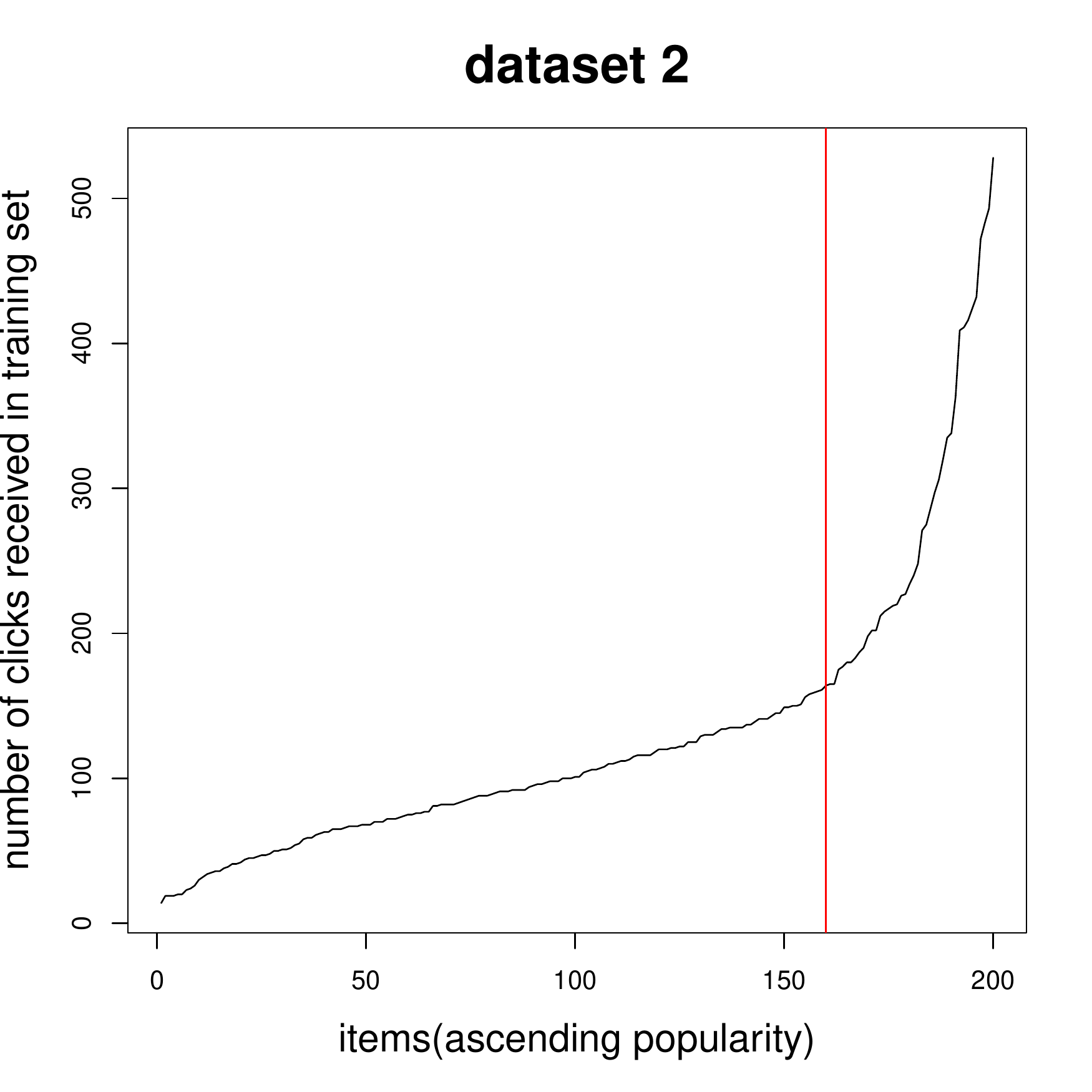}
		\subcaption{Dataset 2 }\label{itemFreq20}
	\end{minipage}  
	
	\caption{Item popularity in the training set. X-axis: item labels, arranged according to ascending item popularity(measured by the number of clicks received in the training set).Y-axis: item popularity}
	\label{fig:itemFreq}
\end{figure}

\begin{table}[!htb]
	\begin{minipage}{.47\linewidth}
		\centering
		\caption{Comparison of rare items recommendations and number of users with at least 1 rare recommendation }
		\label{table:rareRecs}
		\resizebox{\textwidth}{!}{
			\begin{tabular}{ccc}
				\hline\hline
				Dataset&\# rare recs& \# users w. $\geq$ 1 rare recs\\
				\hline
				\multicolumn{1}{c}{\multirow{2}{*}{Dataset 1}}&CF:4388 (5.6\%) &CF:929 (11.8\%)\\
				\multicolumn{1}{c}{}&BMCD: 10071 (12.8\%)& {BMCD: 3454 (43.9\%)}\\
				\hline
				\multicolumn{1}{c}{\multirow{2}{*}{Dataset 2}}&CF: 171 (4.0\%)  &CF: 58 (2.7\%)\\
				\multicolumn{1}{c}{}&BMCD:8770 (20.5\%) &BMCD: 1667 (77.8\%)\\
				\hline
				
		\end{tabular}}
	\end{minipage}%
	\hfill
	\begin{minipage}{.47\linewidth}
		\centering
		\caption{Comparison of intra-list similarity and novelty }
		\resizebox{\textwidth}{!}{
			\begin{tabular}{ccc}
				\hline\hline
				Dataset&intra-list similarity& novelty\\
				\hline
				\multicolumn{1}{c}{\multirow{2}{*}{Dataset 1}}&CF:11.78 &CF:5.97\\
				\multicolumn{1}{c}{}&BMCD: 10.75& {BMCD: 6.18}\\
				\hline
				\multicolumn{1}{c}{\multirow{2}{*}{Dataset 2}}&CF:42.69  &CF: 6.31\\
				\multicolumn{1}{c}{}&BMCD:39.10 &BMCD: 6.60\\
				\hline
			\label{table:intraSim_novelty}		
		\end{tabular}}
	\end{minipage} 
\end{table}

A comparison of intra-list similarity and novelty is shown in Table {\ref{table:intraSim_novelty}}. Consistent with the simulation, BMCD recommends to each user a list of more diverse items, obtaining a lower intra-list similarity score compared to CF. At the same time, BMCD has a stronger ability to recommend more rare and novel items to the users.

% It is also noteworthy to point out that the Mallows method recommends many more ``rare'' items to the users. In \textbf{Table \ref{table:rareItems}}, it displays the number of recommendations that are attributed by less popular items. For dataset 1, the ``less popular'' items refer to those that received fewer than 300 clicks, and for dataset 2, it refers to the items that are clicked by 100 or fewer users. 

% \begin{table}[h!]
% 	\centering
% 	\begin{tabular}{||c c c ||} 
% 		\hline
% 		Method& Dataset 1 & Dataset 2 \\ [0.5ex] 
% 		\hline\hline 
% 		Mallows & 2339&  1744 \\ 
% 		CF & 1540 &23 \\ 
% 		\hline
% 	\end{tabular}
% 	\caption{number of recommendations of rare items }
% 	\label{table:rareItems}
% \end{table}

% \subsubsection{Serendipity}
%A comparison of unserendipity is shown in \textbf{Table \ref{table:serendipity}}. Neither Mallows or CF's recommendations are very serendipitious. This is in part, due to the dataset consisting only 200 items, and the fact that users preferences are quite similar to the others. Overall, Mallows' method shows a very slight advantage. 
%\begin{table}[h!]
%	\centering
%	\begin{tabular}{||c c c ||} 
%		\hline
%		Method& Dataset 1 & Dataset 2 \\ [0.5ex] 
%		\hline\hline 
%		Mallows & {0.018}&  0.0084 \\ 
%		CF & 0.019 &0.0086 \\ 
%		\hline
%	\end{tabular}
%	\caption{unserendipity }
%	\label{table:serendipity}
%\end{table}

\section{Further discussions and future works \label{sec:summary}}

%\subsubsection{Summary and discussions on recommendation diversity}
In this paper, we have introduced and applied BMCD to make personalized recommendations based on clicking data. We have also compared the recommendation performances of BMCD with the popular Collaborative Filtering, in terms of accuracy and diversity. 

Through a simulation study and an offline testing of a dataset, we have observed that BMCD and CF make recommendations with similar level of accuracy. BMCD, in addition, produces interpretable uncertainty estimation for each recommendation made. We showed that the uncertainty can be further exploited to improve the overall accuracy.

We have also assessed the recommendation diversity of both methods through measures of coverage, correct coverage, intra-list similarity and novelty. We have found that compared to CF, BMCD has stronger ability to recommend diverse and rare items to the users, and considers more items for recommendations. 

There are several reasons that explain this phenomenon. First, BMCD, by construction, follows the restriction that all items clicked by a user, need to be among the user's top-ranked items, regardless of how unusual the clicked items are.   This restriction enforces every user's uniqueness, and helps capture and preserve each individual user's ``peculiar'' behavior. CF on the other hand, often sacrifices the ``unusual'' items in the matrix factorization process, since the unusual items contribute less to the cost function. 

Second, {BMCD is sensitive to the clicks in the sparse part of the dataset}. BMCD contains the consensus parameter $\bm{\rho}$, and for the highly sparse part of the dataset, when an item receives a few clicks, these clicks will impact the distribution of the consensus parameter $\bm{\rho}$, and even more so, certain summary statistics such as the Maximum A Posteriori. The consensus, in turn, has an impact on the distribution and summary statistics of the individual users' latent full ranking vectors $\tilde{\bm{{R}}}_j$. However, it can also be a double-edge sword: when the sparse information turns out to be inaccurate or unrepresentative, it can decrease the method's recommendation accuracy.

It is therefore not surprising to observe that recommendations using BMCD are often more diverse, involving more rare items, even when user behaviors are rather homogeneous. BMCD's strong ability to capture the perculiarity of the users and its tendency to recommend less popular items partly explains why it was marginally outperformed in terms of accuracy by CF in the offline testing scenario, where the ground truth is limited by what the users have already seen and clicked. Rare items are often not yet discovered by the users, and it is almost impossible to verify the success of such recommendations in an offline testing. We are currenlty planning online testing of BMCD.
%also hinders its ability to exploit items' popularity, which in a great extent, explains why CF marginally beats BMCD in terms of recommendation accuracy in certain cases. 

%There are many ways to further enhance the recommendation performance of BMCD. In this study, we have considered the situation where users' preference data is in the form of binary clicks, which we assumed that a clicked item is more preferrred than the unclicked items. Apart from the clicks, other information can also be incorporated into the model, such as the time sequence of the clicks. When presented together, an item that is clicked first is likely to be more preferred by the user compared to an item clicked later. Other forms of preference data, such as ratings and thumbs up/down can also be incorporated into the model since they all can be easily transformed into partial rankings. 

One of the biggest drawbacks of BMCD, which is based on MCMC, is scaling. BMCD does not scale well due to the huge amount of parameters to be estimated.{ The computing time required is dependent on the number of users $N$, the number of clusters $C$, as well as the number of iterations required to reach convergence. It takes 53 hours to compute for 1 million iterations for the NRK dataset 1, and 14 hours for dataset 2 using one core of the Intel Xeon e-8890 processor, running at 2.5 GHz.} The iterative nature of MCMC also makes efficient parallelization more challenging since the computational overhead is very heavy. {The Spark implementation of CF on the other hand, is very efficient. However, it can also be computationally costly if a thorough cross-validation is to be performed}. For BMCD, in practice, it often happens that the cluster assignments for each user, $z_1, ..., z_N$ converge quite quickly. In the situation that most users do not switch cluster memberships often, after the cluster assignments have converged, we can split the dataset into $C$ different segments, and compute BMCD algorithm without the clustering steps independently and in parallel. The reduction in the number of users, and the number of parameters needed to be estimated, {can reduce computing time to at least 1/C of its original required computing time if the $C$ clusters are similar in size}. Another way to speed up the computation is by choosing smart starting points for the MCMC such that convergence can be reached in fewer iterations. We suggest, for example, that instead of randomly initializing the cluster assignment for each user, $z_1, ..., z_N$ can be initialized based on a K-means clustering. We are currently working on variational Monte carlo versions of our algorithm, which we expect to reduce computational time very significantly.

In conclusion, BMCD can be considered as a valid alternative to traditional state-of-the-art collaborative filtering when diversity in the recommendation is an important objective.
\section{Acknowledgement}
We give special thanks to Linn Cecilie Solbergersen and the Norwegian Broadcasting Company for generously providing us research data and kind collaborations. We also thank \O ystein S\o rensen, Elja Arjas  and Valeria Vitelli for fruitful discussions.
\newpage

\section*{References}
\bibliographystyle{elsarticle-num}
\bibliography{references_kbs}   

%\begin{appendix}

%\section{BMCD MCMC algorithm}\label{sec:algo}

\end{document}

\maketitle

\section{Boxplots of within cluster sum of footrule distance for the simulation study\label{sec:simWithinSS}}
The three figures below show the distributions of the sum of  within-cluster sum of footrule distance, plotted against the various numbers of clusters $C$, for 3 selected runs. It can be observed from the figures, $C = 3$ is chosen for run number 1 and 10, while $C = 4$ is chosen for run number 5. 
\begin{figure}[htb!]
	\begin{minipage}[t]{.31\textwidth}
		\centering
		\includegraphics[width=\textwidth]{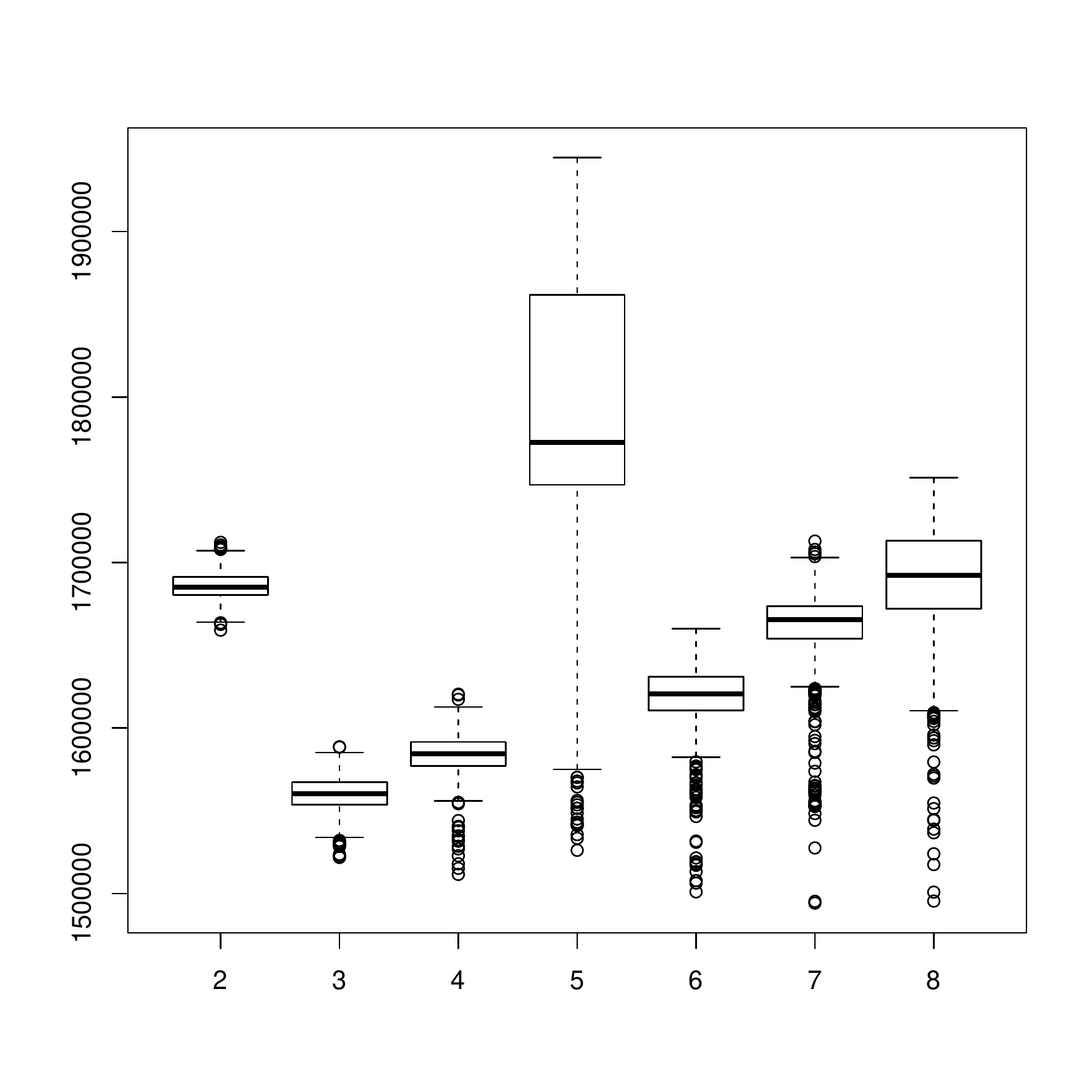}
		\subcaption{Run 1}\label{wcd:1}
	\end{minipage}
	\hfill
	\begin{minipage}[t]{.31\textwidth}
		\centering
		\includegraphics[width=\textwidth]{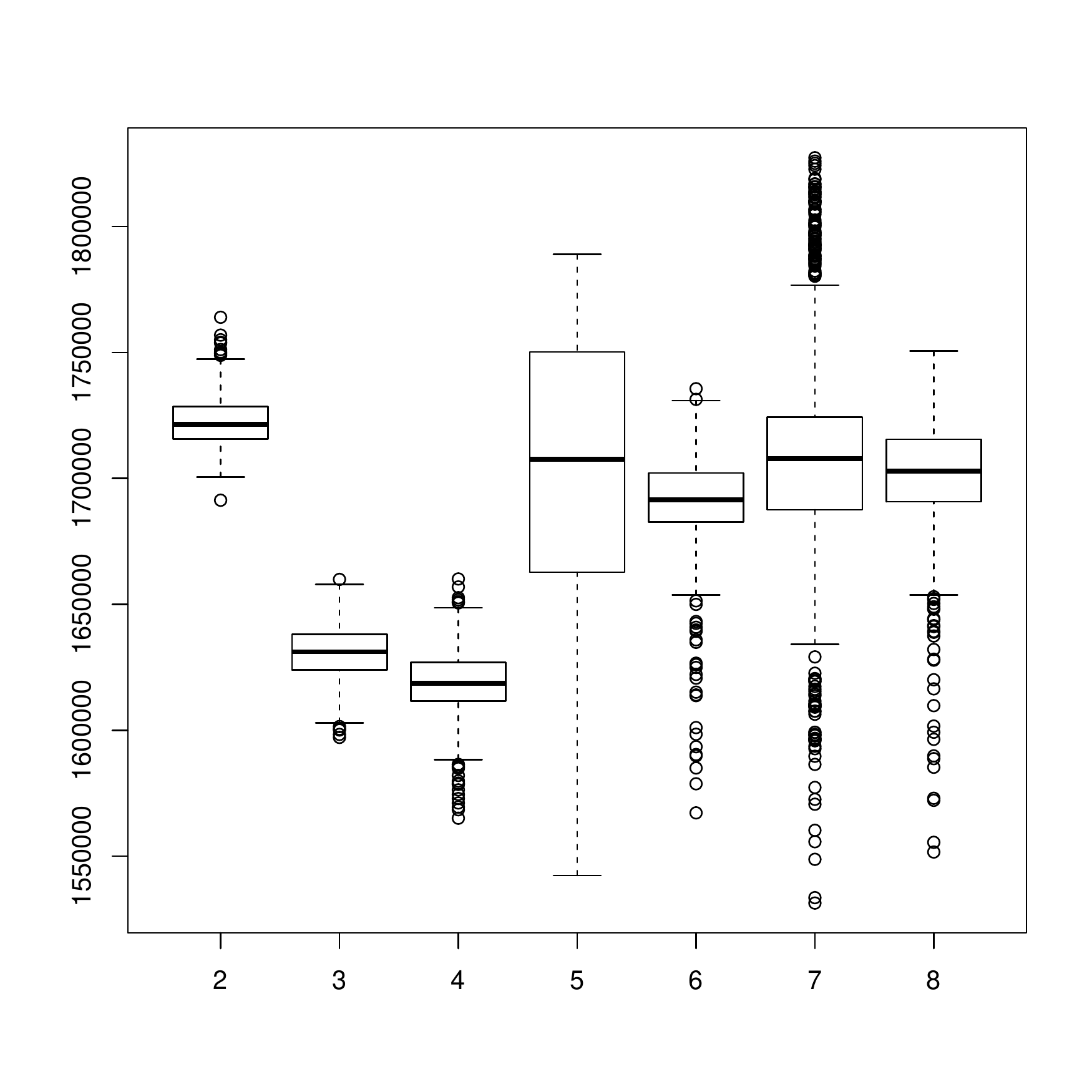}
		\subcaption{Run 5}\label{wcd:2}
	\end{minipage}  
	\hfill
	\begin{minipage}[t]{.31\textwidth}
		\centering
		\includegraphics[width=\textwidth]{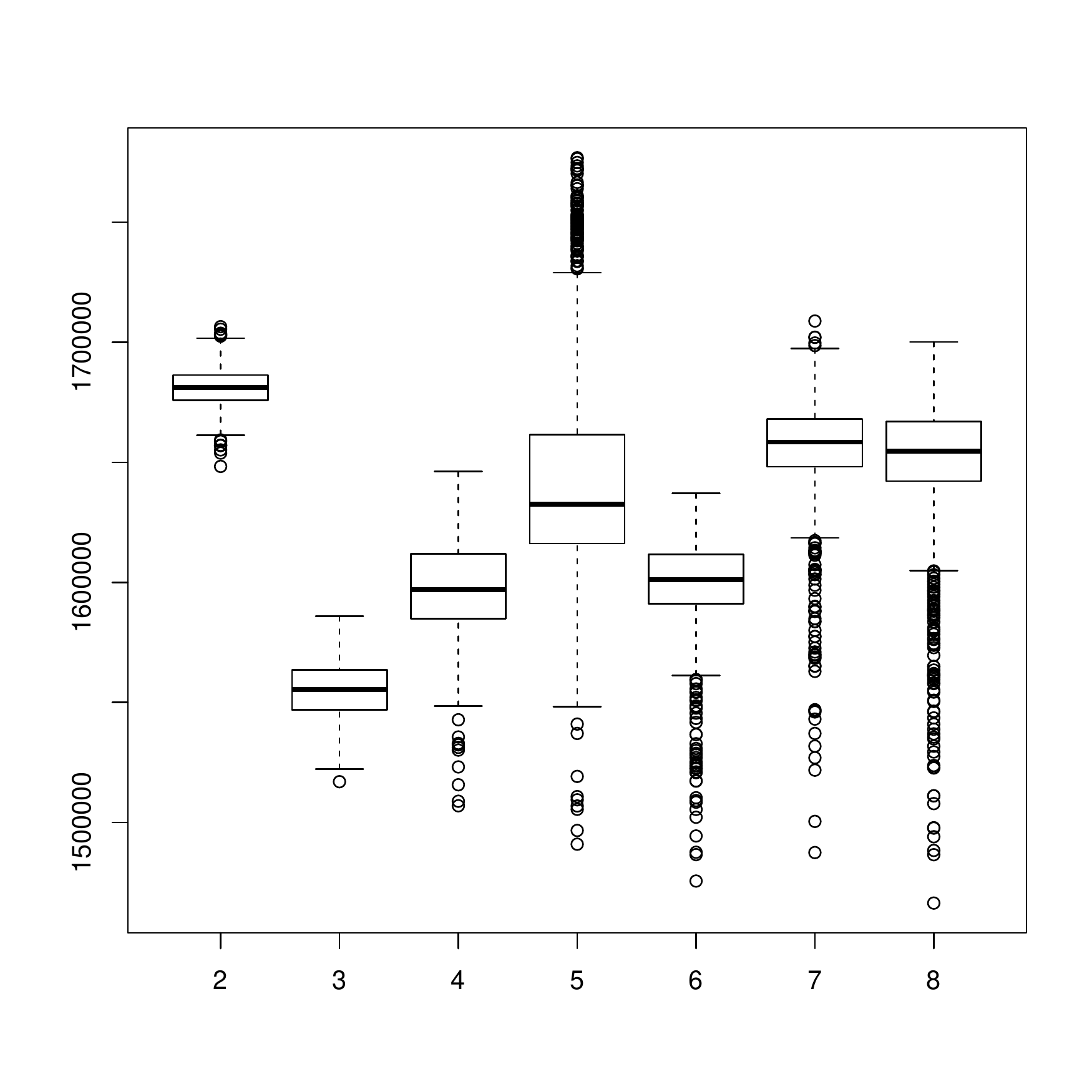}
		\subcaption{Run 10}\label{wcd:3}
	\end{minipage}  
	\caption{Posterior sum of within cluster distances of selected runs}
	\label{fig:wcd}
\end{figure}

\section{Traceplots of $\alpha_c$ of simulation study
	\label{sec:convergenceSim}}
The three figures shown below are the traces of $\alpha_1, \alpha_2$, and $\alpha_3$ after the burn-in period, for a selected run. It can be seen that the posterior mean is roughly 3 for all 3 clusters, reflecting the true value of $\alpha$.
\begin{figure}[htb!]
	\begin{minipage}[t]{.31\textwidth}
		\centering
		\includegraphics[width=\textwidth]{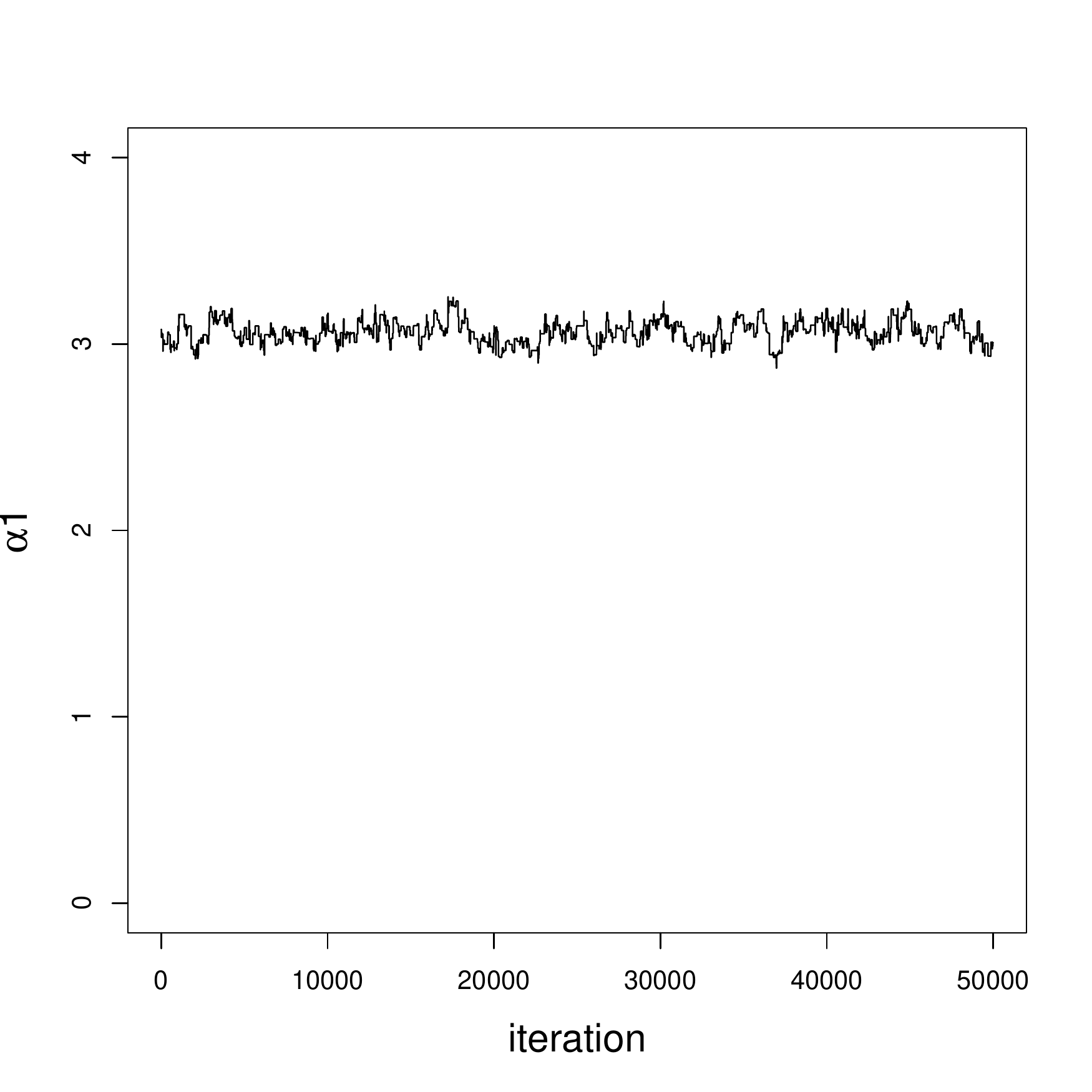}
		\subcaption{Run 1, $\alpha_1$}\label{alpha:1}
	\end{minipage}
	\hfill
	\begin{minipage}[t]{.31\textwidth}
		\centering
		\includegraphics[width=\textwidth]{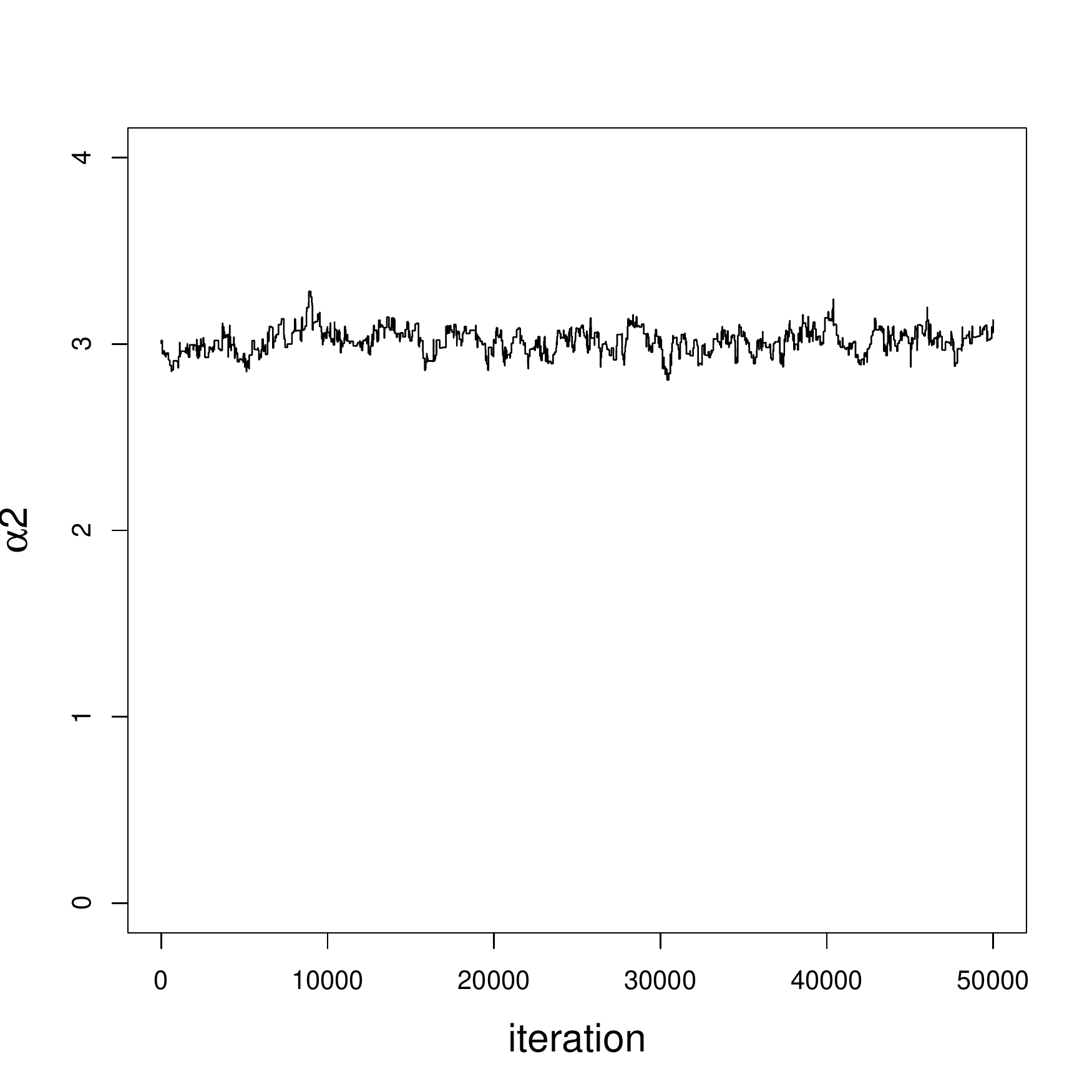}
		\subcaption{Run 1,  $\alpha_2$}\label{alpha:2}
	\end{minipage}  
	\hfill
	\begin{minipage}[t]{.31\textwidth}
		\centering
		\includegraphics[width=\textwidth]{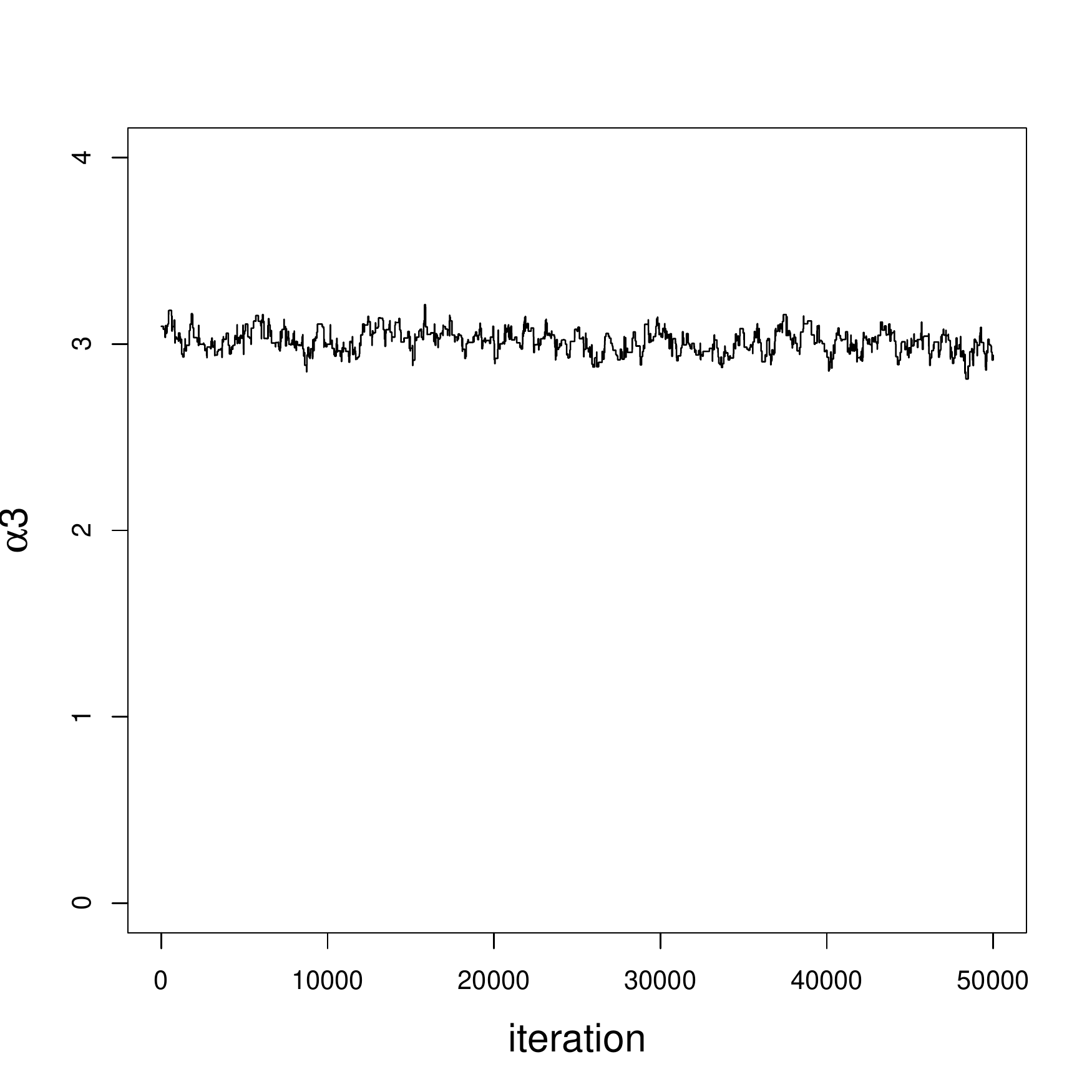}
		\subcaption{Run 1,  $\alpha_3$}\label{alpha:3}
	\end{minipage}  
	\caption{trace of $\alpha_1, \alpha_2, \alpha_3$ after burn-in period, run 1}
	\label{fig:alphaTraceSim}
\end{figure}

\section{Traceplots of $\alpha_c$ for the NRK datasets\label{sec:convergenceNRK}}
The figures below show the traces of $\alpha$ values after the burn-in period for 3 selected clusters. It can be observed that cluster 1 of dataset 1 has a higher value of $\alpha$, suggesting that users in this cluster have more similar preferences. 
\begin{figure}[htb!]
	\begin{minipage}[t]{.31\textwidth}
		\centering
		\includegraphics[width=\textwidth]{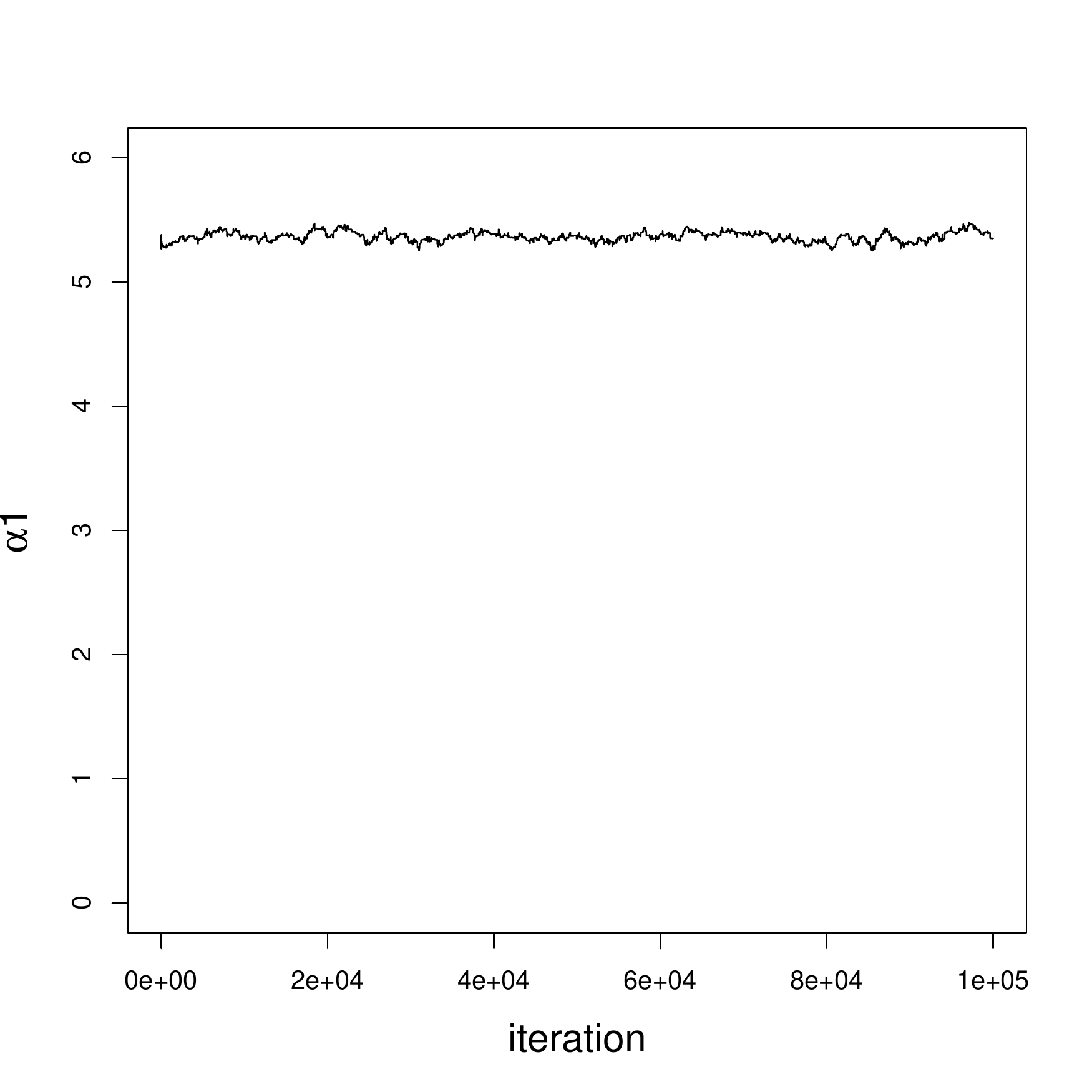}		
		\subcaption{Dataset 1, cluster 1}\label{nrk10_alpha_1}
	\end{minipage}
	\hfill
	\begin{minipage}[t]{.31\textwidth}
		\centering
		\includegraphics[width=\textwidth]{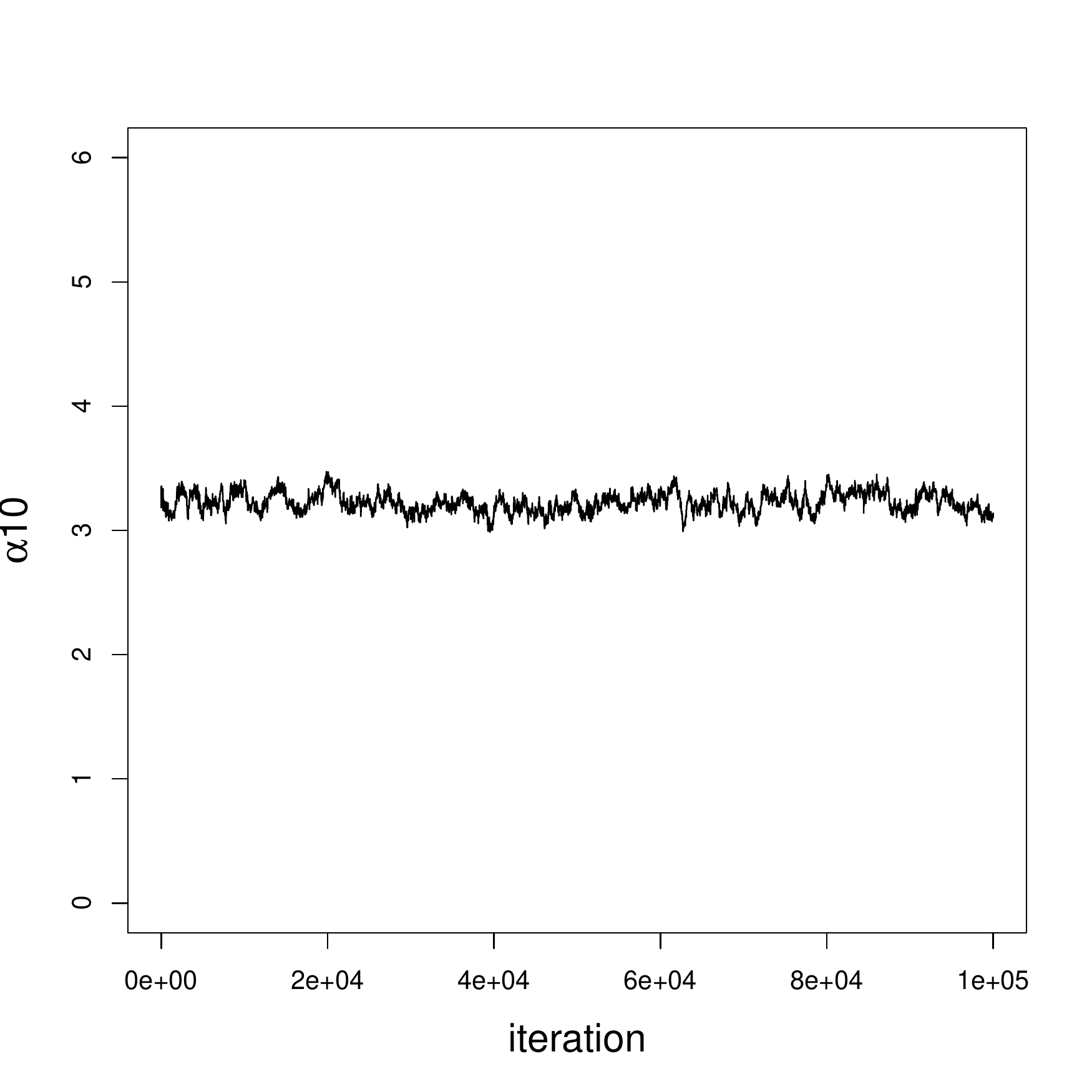}
		\subcaption{Dataset 1, cluster 10}\label{nrk10_alpha_10}
	\end{minipage}  
	\begin{minipage}[t]{.31\textwidth}
		\centering
		\includegraphics[width=\textwidth]{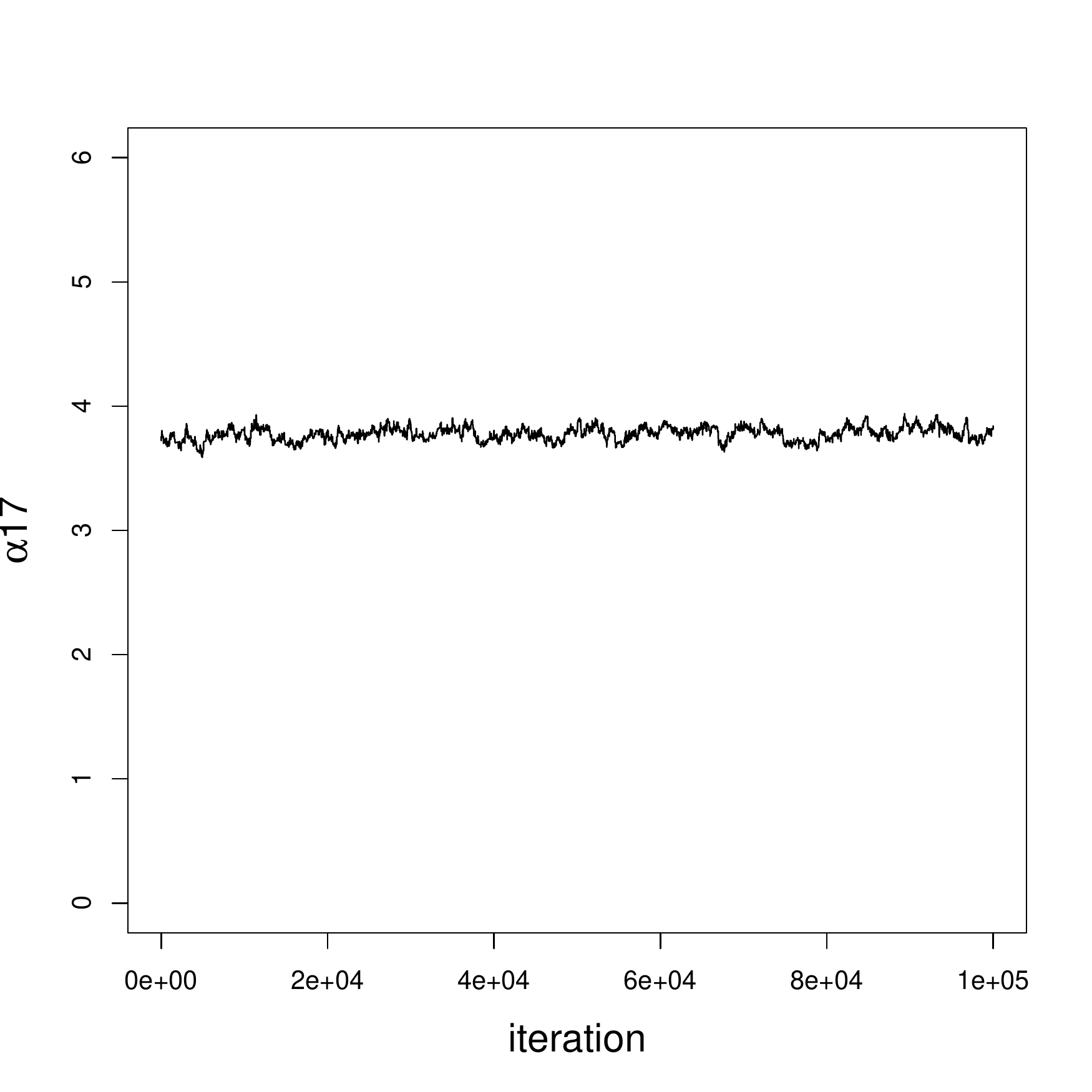}
		\subcaption{Dataset 1, cluster 17}\label{nrk10_alpha_17}
	\end{minipage} 
	\begin{minipage}[t]{.31\textwidth}
		\centering
		\includegraphics[width=\textwidth]{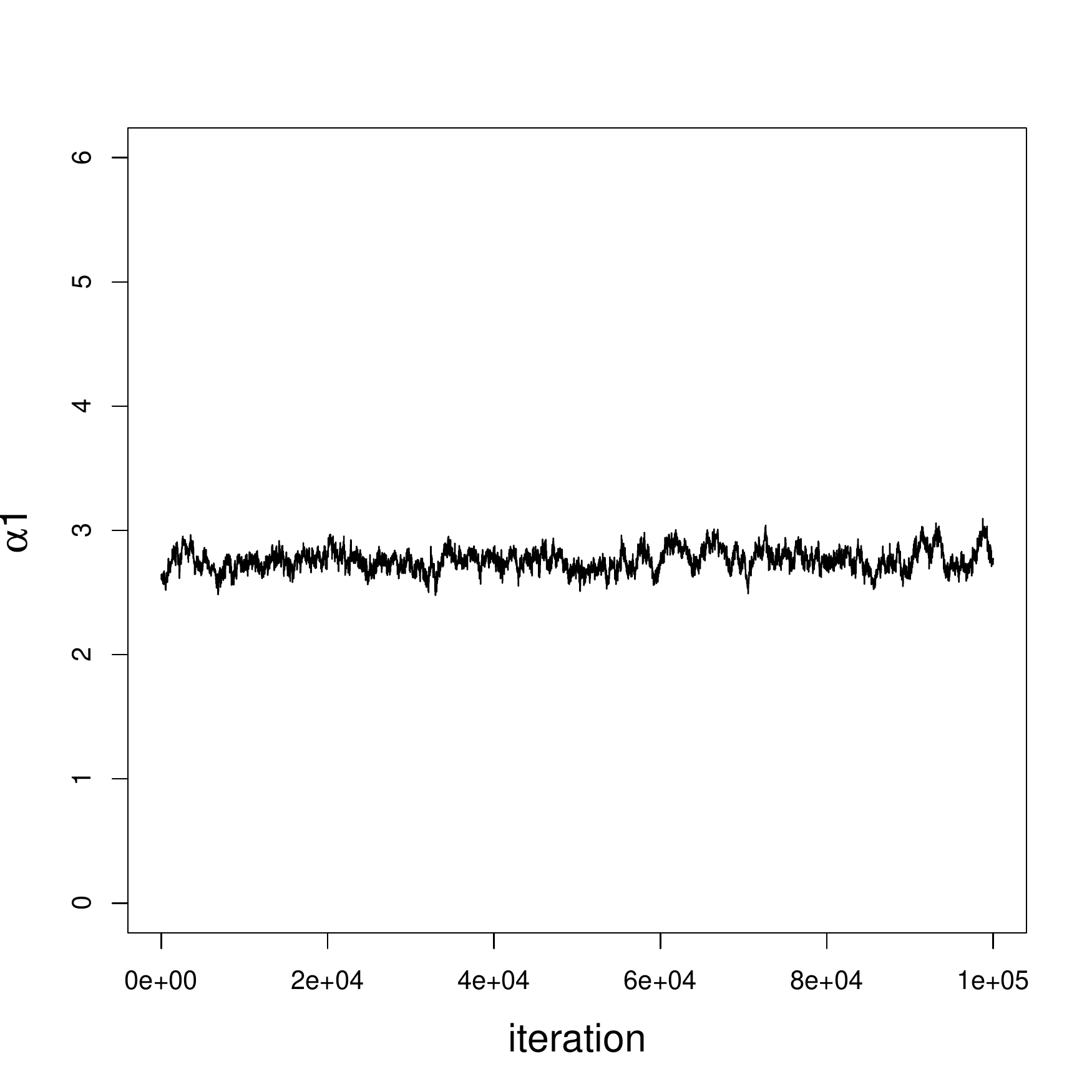}
		\subcaption{Dataset 2, cluster 1}\label{nrk20_alpha_1}
	\end{minipage}
	\hfill
	\begin{minipage}[t]{.31\textwidth}
		\centering
		\includegraphics[width=\textwidth]{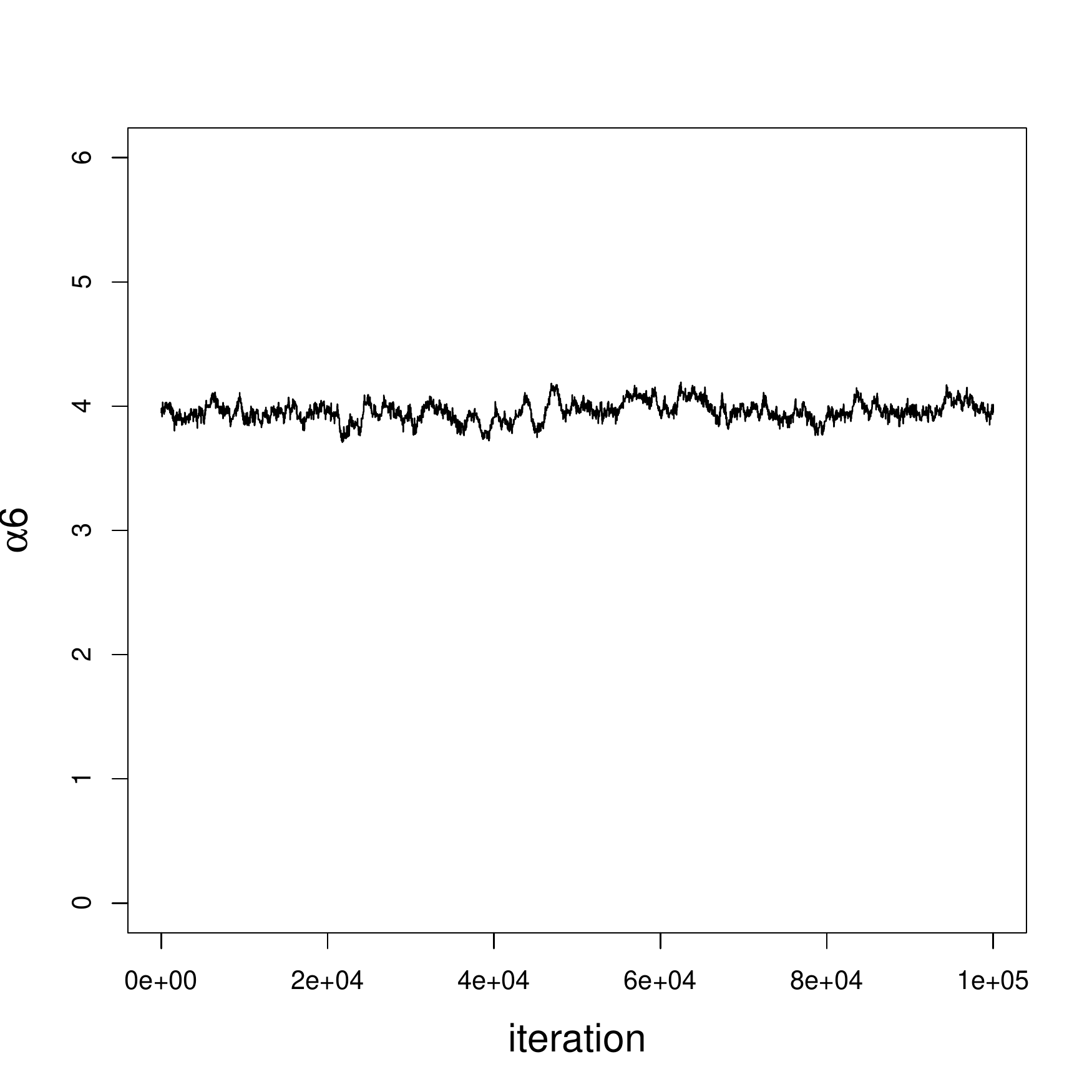}
		\subcaption{Dataset 2, cluster 6}\label{nrk20_alpha_6}
	\end{minipage}  
	\begin{minipage}[t]{.31\textwidth}
		\centering
		\includegraphics[width=\textwidth]{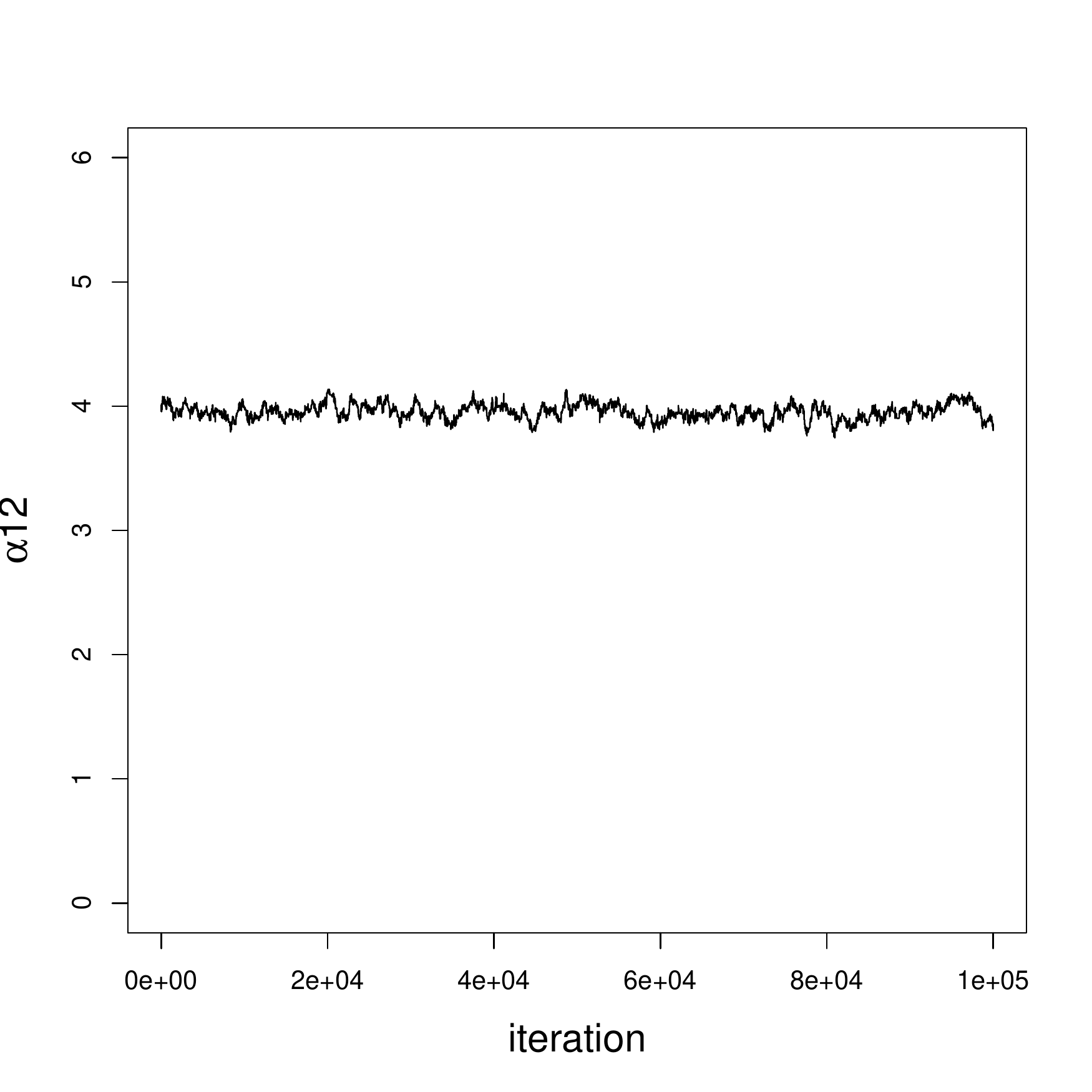}
		\subcaption{Dataset 2, cluster 12}\label{nrk20_alpha_12}
	\end{minipage} 
	\caption{Traces of $\alpha_c$ after the burn-in period for convergence check}
	\label{fig:NRK_alpha}
\end{figure}

\section{Kmeans within-cluster sum of square for the NRK datasets}
The figures below show the within-cluster sum of square plotted against the varying number of clusters $C$. It is possible to notice the change in gradient, but the elbow is not very obvious. Combining the elbow and a preference towards larger number of clusters, $C=17$ and $C=12$ is chosen, indicated by the red dotted line, for dataset 1 and 2 respectively.

\begin{figure}[h!]
	\begin{minipage}[t]{.35\textwidth}
		\centering
		\includegraphics[width=\textwidth]{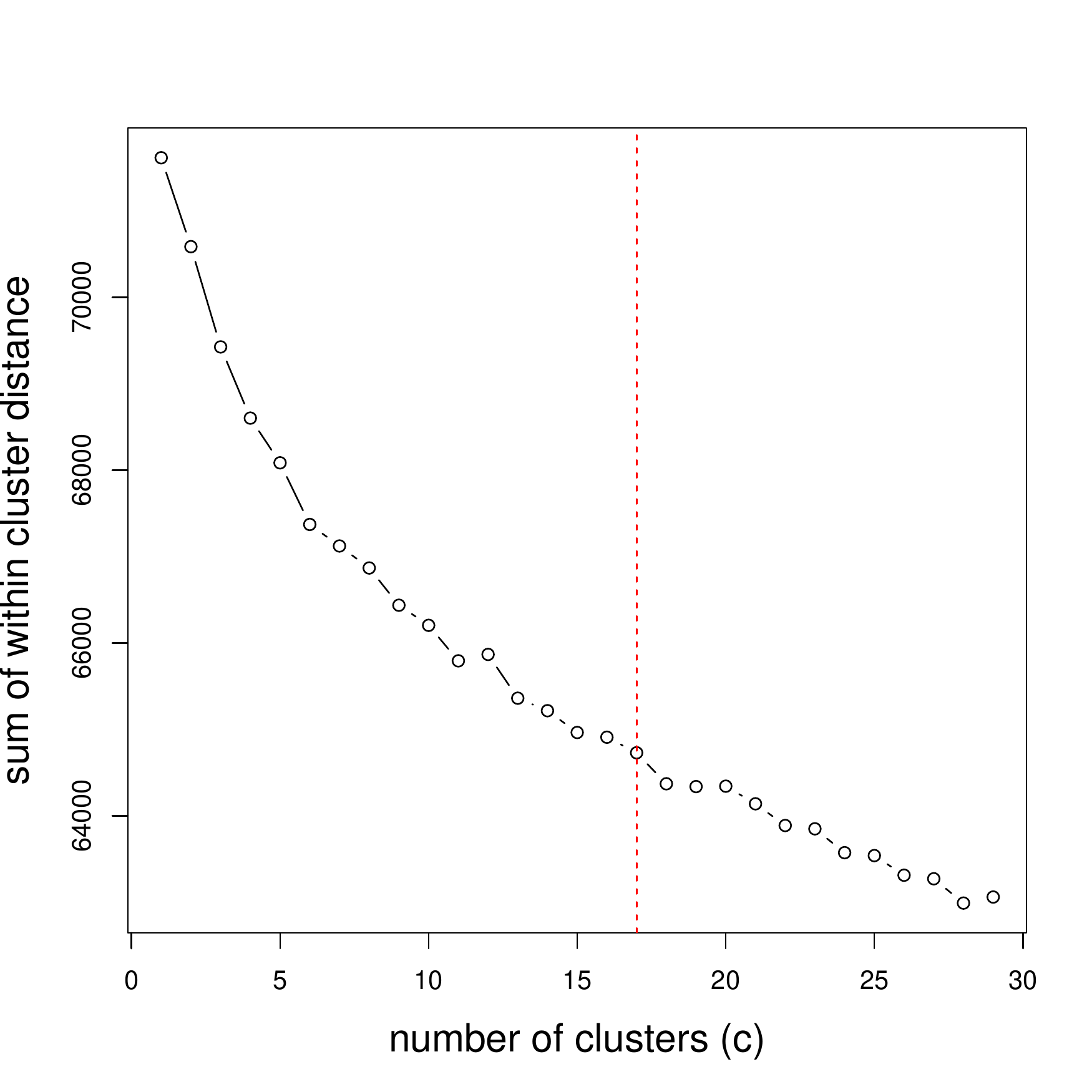}
		\subcaption{Dataset 1}\label{C_determine_1}
	\end{minipage}
	\hfill
	\begin{minipage}[t]{.35\textwidth}
		\centering
		\includegraphics[width=\textwidth]{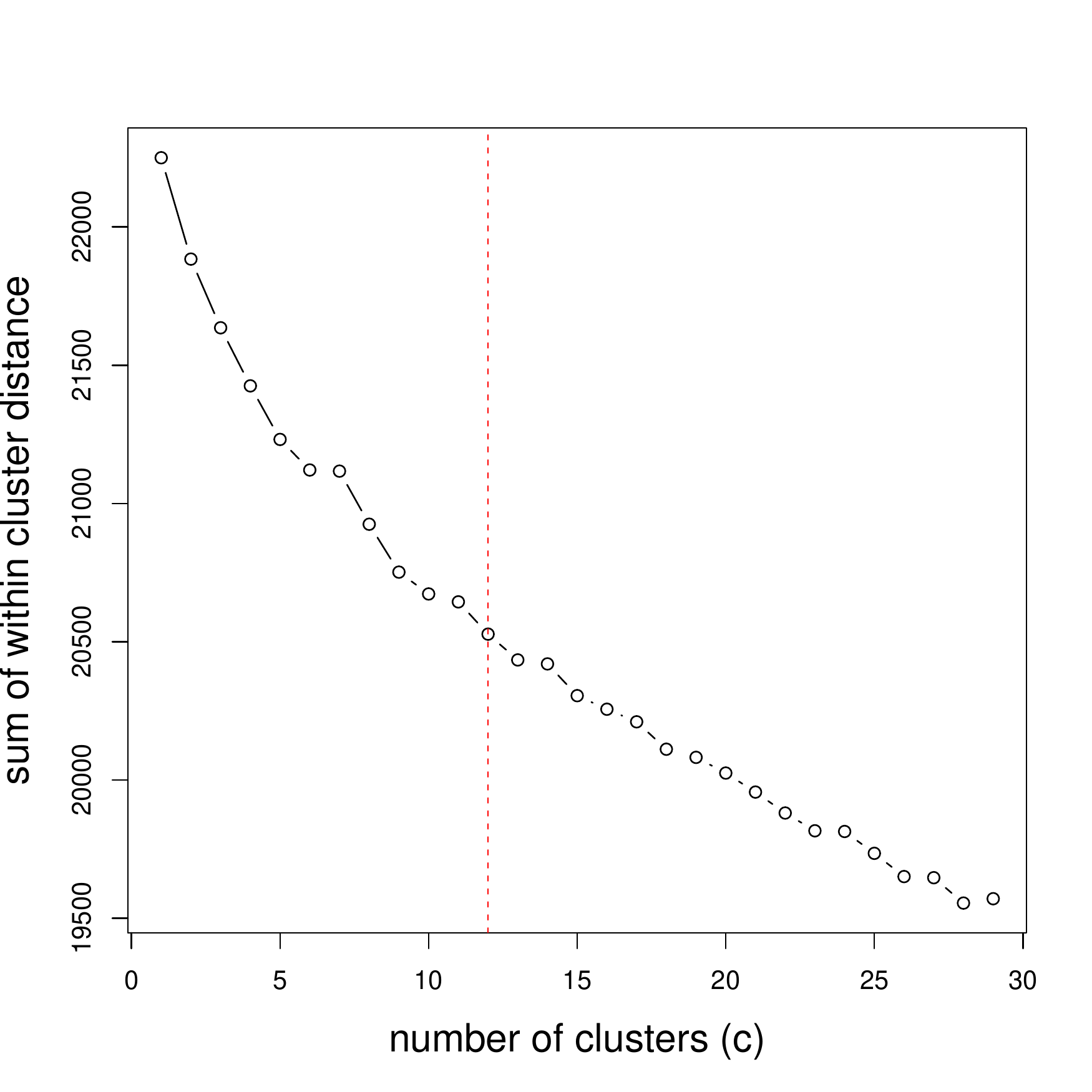}
		\subcaption{Dataset 2}\label{C_determine_2}
	\end{minipage}  
	
	\caption{Within-cluster sum of square using K-means clustering vs. number of clusters chosen}
	\label{fig:Kmeans}
\end{figure}

\newpage
\section[h!]{Algorithms \label{sec:algo}}
In this section, we present the BMCD algorithms, including the initialization process, MCMC estimation, and the recommendation procedure.

\begin{algorithm}[h!]
	\SetAlgoLined
	\DontPrintSemicolon
	\KwIn{$C$, \{$\mathcal{A}_1, ..., \mathcal{A}_N$\}, $n$, $N$, \{$c_1, ..., c_N$\}}
	\KwOut{\{$z_1^0,...,z_N^0$\}, \{$\alpha_1^0,..., \alpha_C^0$\}, \{$\bm{\rho}_1^0, ..., \bm{\rho}_C^0$\}, \{$\tilde{\bm{R}}_1^0,...,\tilde{\bm{R}}_N^0$\}},	\{$\tau_1^0, ..., \tau_C^0$\}
	randomly generate	\{$\tau_1^0, ..., \tau_C^0$\} \;
	\eIf{random initialization}{
		randomly generate \{$z_1^0, ..., z_N^0$\} $\in \{1, ..., C\}$, \{$\alpha_1^0, ..., \alpha_C^0\} \in (0, \infty)$, $\{\bm{\rho}_1^0, ..., \bm{\rho}_C^0\} \in$  $\mathcal{P}_n$\;
		%\For{c $\gets$ 1 to C}{
		%		randomly draw  $\bm{\rho}_c^0$ from $\mathcal{P}_n$
		%		}
		
		\For{j $\gets$ 1 to N}{
			\For{i $\in$ $\mathcal{A}_j$}{
				randomly draw $R_{ij}^0$ from $\mathcal{P}_{c_j}$			
			}
			\For{i $\in$ $\mathcal{A}_j^c$}{
				randomly draw $R_{ij}^0$ from $\mathcal{P}_{n-c_j}$	 \;
				$R_{ij}^0 \gets R_{ij}^0 + c_j $
			}
			
		}
		
	}{
		\For{$i \gets 1$ to $n$}{
			compute item frequency $Freq_i$ based on \{$\mathcal{A}_1, ..., \mathcal{A}_N$\}\;
		}
		generate $\bm{\rho}^0$ based on $Freq_1, ..., Freq_n$
		
		\For{$j \gets 1$ to $N$}{
			initialize $\tilde{\bm{R}}_j^0$	based on $\mathcal{A}_j$ and $\bm{\rho}^0$
		}
		\For{$j \gets 1$ to $N$}{
			randomly generate $z_j^0 \in \{1, ..., C\}$
		}
		\For{$c \gets 1$ to $C$}{
			\For{$i \gets 1$ to $n$}{
				compute $Freq_{c,i}$ based on \{$\mathcal{A}_j: z_j =c$\}
			}
			compute $\bm{\rho}_c^0$ based on \{$Freq_{c,1}$,...,$Freq_{c,n}$\}
		}

	}
	
	\caption{MCMC initialization}
	\label{algo:init}
\end{algorithm}
\newpage
\begin{algorithm}[H]	
	\SetAlgoLined
	\DontPrintSemicolon
	\KwIn{iter.max,  $\alpha_{update}$, \{$\tau_1^0, ..., \tau_C^0$\}, \{$z_1^0,...,z_N^0$\}, \{$\alpha_1^0,..., \alpha_C^0$\}, \{$\bm{\rho}_1^0, ..., \bm{\rho}_C^0$\}}
	\KwOut{ $\{\tilde{\bm{R}}_1,...,\tilde{\bm{R}}_N\}^{1, ..., \text{iter.max}}$, $\{z_1, ..., z_N\}^{1,..., \text{iter.max}}$, $\{\bm{\rho}_1, ..., \bm{\rho}_C\}^{1,..., \text{iter.max}}$, $\{\alpha_1, ..., \alpha_C\}^{1,...,\text{iter.max}/\alpha_{update}}$}
	
	\For{$t \gets 1$ to \text{iter.max}}{
		\textbf{Gibbs} update $\{\tau_1,..., \tau_C\}^{t} \gets \{\tau_1,..., \tau_C\}^{t-1}$ \;
		\For{$c \gets$ 1 to $C$}{
			\textbf{M-H} update $\bm{\rho}_c^t \gets \bm{\rho}_c^{t-1}$ \;
			\If{$t$ mod $\alpha_{update}$ ==0}{
				\textbf{M-H} update ${\alpha}_c^t \gets {\alpha}_c^{t-1}$
			}
		}
		\For{$j \gets$ 1 to $N$ \textbf{in parallel}}{
			\textbf{Gibbs} update $z_j^t \gets z_j^{t-1}$ \;
			\textbf{M-H} update $\tilde{\bm{R}}_j^{t} \gets \tilde{\bm{R}}_j^{t-1}$
			
		}
	}
	
	\caption{MCMC algorithm}
	\label{algo:MCMC}
\end{algorithm}

\begin{algorithm}
	\SetAlgoLined
	\DontPrintSemicolon
	\KwIn{$\{\mathcal{A}_1,...,\mathcal{A}_N\}, \{c_1, ..., c_N\}, \{\tilde{\bm{R}}_1,...,\tilde{\bm{R}}_N\}^{1, ..., \text{iter.max}}$, $k$}
	\KwOut{$\{Rec_{11}, ..., Rec_{k1}\},...,\{ Rec_{1N}, ..., Rec_{kN}\}$}
	\For{j $\gets$ 1 to N}{
		\For{i $\gets$ 1 to n}{
			compute $P_{ij} = P(\tilde{R}_{ij}<c_j + k | \mathcal{A}_1,...,\mathcal{A}_N)$
		}
	}
	
	\For{j $\gets$ 1 to N}{
		ranked($\bm{P}_{j} ) = \{P_{j(1)}, ...,P_{j(n)} \}$	 \;
		Recommend $Rec_{j,k} = \{A_r: P_{rj} \in \{P_{j(1)}, ...,P_{j(k)}$\} \}}
	
	\;

	\caption{Making recommendations}
	\label{algo:postproc}
\end{algorithm}